\documentclass[aps,prb,twocolumn,superscriptaddress,fleqn]{revtex4-1}
\usepackage{graphicx}
\usepackage[version=3]{mhchem}
\usepackage{chemfig,wrapfig}
\usepackage{float}
\usepackage{siunitx}
\usepackage[hidelinks]{hyperref}
\usepackage{color}
\usepackage{xspace}

\newcommand{\lsco} {{La$_{2-x}$Sr$_x$CuO$_4$}\@\xspace}

\newcommand{\ybcoF} {$\ce{YBa2Cu3O_{7}}$\@\xspace}
\newcommand{\ybco} {$\ce{YBa2Cu3O_{6+y}}$\@\xspace}

\newcommand{\ybcoE} {$\ce{YBa2Cu4O8}$\@\xspace}

\newcommand{\tc} {\ensuremath{T_c}\@\xspace}
\newcommand{\temp} {\ensuremath{T}\@\xspace}

\newcommand{\cpara}[0]{\ensuremath{{{c\parallel B_0}}}\@\xspace}
\newcommand{\cperp}[0]{\ensuremath{{c\perp B_0}}\@\xspace}

\begin{document}

\title{Phenomenology of $^{63}$Cu nuclear relaxation in cuprate superconductors}

\author{Michael Jurkutat}
\affiliation{University of Leipzig, Felix Bloch Institute for Solid State Physics,  
Linnestr. 5, 04103 Leipzig, Germany}
\author{Marija Avramovska}
\affiliation{University of Leipzig, Felix Bloch Institute for Solid State Physics,  
Linnestr. 5, 04103 Leipzig, Germany}
\author{Grant V. M. Williams}
\affiliation{Victoria University of Wellington, The MacDiarmid Institute for Advanced Materials and Nanotechnology, SCPS, PO Box 600, Wellington 6140, New Zealand} 
\author{Daniel Dernbach}
\affiliation{University of Leipzig, Felix Bloch Institute for Solid State Physics,  
Linnestr. 5, 04103 Leipzig, Germany}
\author{Danica Pavi\'cevi\'c}
\affiliation{University of Leipzig, Felix Bloch Institute for Solid State Physics,  
Linnestr. 5, 04103 Leipzig, Germany}
\author{J\"urgen Haase* }
\affiliation{University of Leipzig, Felix Bloch Institute for Solid State Physics,  
Linnestr. 5, 04103 Leipzig, Germany}

\date{\today} 
\begin{abstract}
Nuclear relaxation is an important thermodynamic probe of electronic excitations, in particular in conducting and superconducting systems. Here, an empirical phenomenology based on all available literature data for planar Cu in hole-doped cuprates is developed. It is found that most of the seemingly different relaxation rates among the systems are due to a temperature independent anisotropy that affects the mostly measured $1/T_{1\parallel}$, the rate with an external magnetic field along the crystal $c$-axis, while $1/T_{1\perp}$ is largely independent on doping and material above the critical temperature of superconductivity ($T_c$). This includes very strongly overdoped systems that show Fermi liquid behavior and obey the Korringa law. Below $T_c$ the relaxation rates are similar, as well, if plotted against the reduced temperature $T/T_c$. Thus, planar Cu nuclear relaxation is governed by a simple, dominant mechanism that couples the nuclei with varying anisotropy to a rather ubiquitous bath of electronic excitations that appear Fermi liquid-like irrespective of doping and family. In particular, there is no significant enhancement of the relaxation due to electronic spin fluctuations, different from earlier conclusions. Only the La$_{2-x}$Sr$_x$CuO$_4$ family appears to be an outlier as additional relaxation is present, however, the anisotropy remains temperature independent. Also systems with very low doping levels, for which there is a lack of data, may behave differently.
\end{abstract}



\maketitle

\section{Introduction}
Nuclear relaxation is a fundamental probe in condensed matter physics \cite{Slichter1990}. In conducting materials it is often determined by the electronic excitations, as was predicted \cite{Heitler1936} long before the techniques of nuclear magnetic resonance (NMR) became available. The heat transfer that establishes the (electronic) lattice temperature for a nuclear spin system can be conveniently measured in an external magnetic field, but also in zero field. It is characterized by the spin-lattice relaxation rate $1/T_1$. For Fermi liquids, this scattering of nuclear spins off electrons is proportional to the square of the electronic density of states, to the hyperfine coefficients, but also proportional to the temperature ($T$), i.e., $1/T_1 \propto T$. The famous Korringa relation \cite{Korringa1950} is very useful for simple Fermi liquids since the electronic Pauli susceptibility leads to the NMR Knight shift ($K_{\rm S}$) that is also proportional to the electronic density of states and the hyperfine coefficients, so that $1/(T T_1) = (\gamma_n/\gamma_e)^2 (4\pi {\rm k_B}/\hbar)\cdot \rho K_{\rm S}^2$. The gyromagnetic ratios of the electron ($\gamma_e$) and nucleus under study ($\gamma_n$), as well as ${\rm k_B}$ and $\hbar$ are known, and $\rho$ is introduced to account for slight deviations (e.g. due to electronic correlations). 

Another hallmark relation, of relevance here, comes from the first proof of BCS theory of superconductivity \cite{Bardeen1957}. The relaxation rate $1/T_1$ was shown to disappear below the critical temperature of superconductivity (\tc), but the opening of the gap also led to a coherence peak (Hebel-Slichter peak of NMR \cite{Hebel1957}), both predicted by BCS for singlet pairing.

\begin{figure}
 \centering
 \includegraphics[width=0.5\linewidth]{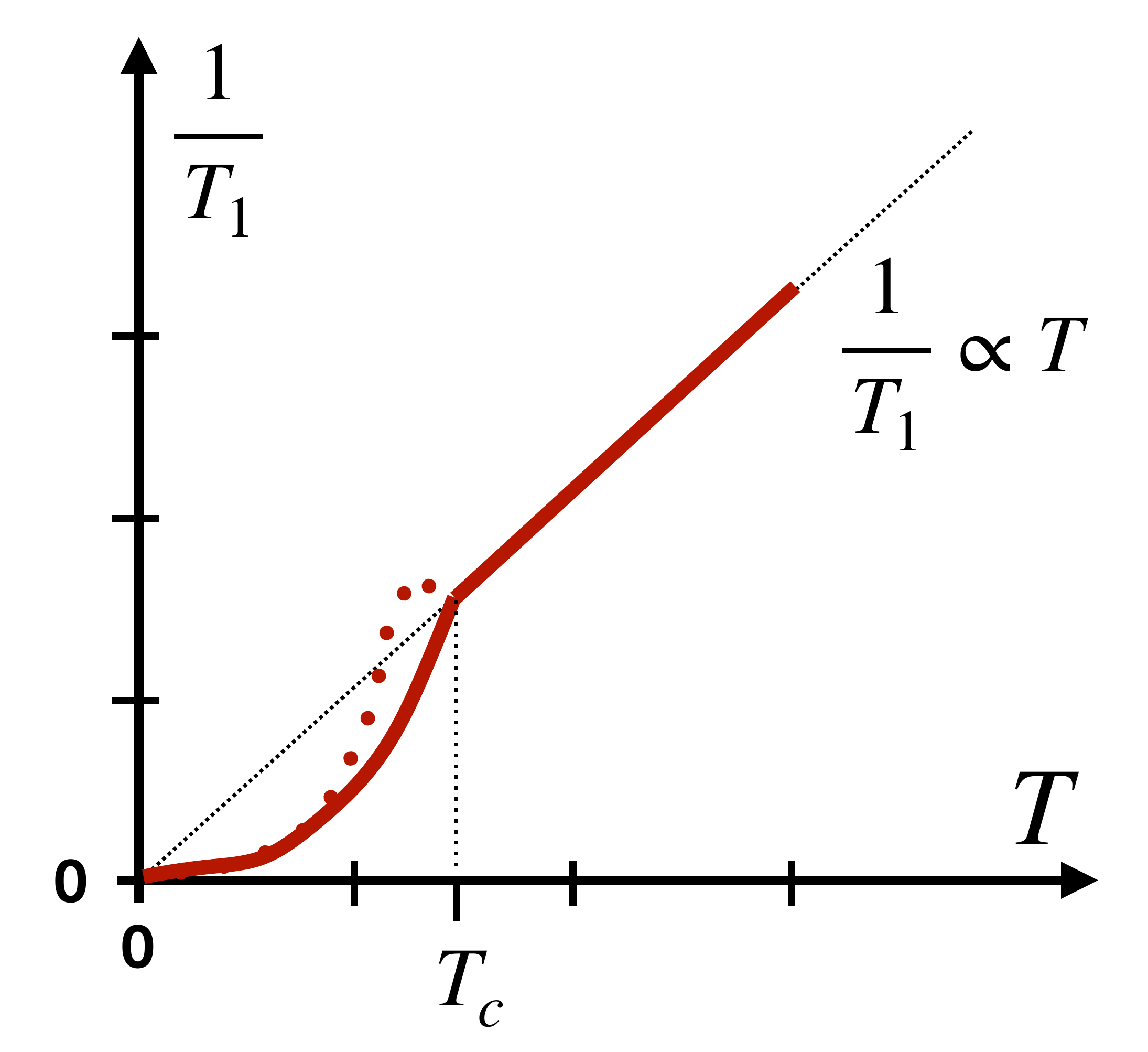}
 \caption{Sketch of the temperature dependence of the nuclear relaxation rate from coupling to a Fermi liquid. The relaxation rate is \emph{proportional} to temperature, above \tc. Below \tc the rate drops for spin singlet pairing and may show a Hebel-Slichter peak just below \tc (dotted line) before it disappears as \temp approaches zero.}
 \label{fig:one}
\end{figure}

Thus, nuclear relaxation, as a bulk sensor of electron thermodynamic properties is a very important probe, and the typical temperature dependence of $1/T_1$ for a classical superconductor is sketched in Fig.~\ref{fig:one}.

With the discovery of cuprate high-temperature superconductors \cite{Bednorz1986} there was immediate interest in measuring the nuclear relaxation rate, in particular for Cu and O nuclei in the ubiquitous \ce{CuO2} plane where the nuclei must couple strongly to the electronic degrees of freedom. Early experiments focussed on the \ybco{} family of materials. The results showed more complicated dependences above and below \tc, even for the apparently overdoped \ybcoF{} that should be closer to a Fermi liquid. Nevertheless, the relaxation did disappear even as $1/(T_1T)$ at low \temp for the latter material in agreement with spin-singlet pairing. The actual decrease of $1/(T_1T)$ as a function of \temp below \tc was weaker than what follows from a symmetric $s$-wave gap. Furthermore, a Hebel-Slichter coherence peak could not be found (it can can also be absent\cite{Kotegawa2001} or broadened\cite{Silbernagle1967} for non-cuprate superconductors).

While relaxation measurements are quite robust, they can be difficult in the cuprates. The large unit cell gives rise to various resonances, and the $^{63,65}$Cu and $^{17}$O nuclei have quadrupole moments. Not only does this lead to even more resonances from angular dependent splittings, one also finds large line broadenings that prove extensive variations of the local electric field gradient (EFG) at Cu and O nuclei, which  are in fact charge density variations as proven more recently \cite{Rybicki2009, Jurkutat2013, Reichardt2018}. In addition, there are many nuclear reservoirs and quadrupolar relaxation could be present, as well. Also, broad resonances that cannot be equally excited with radio frequency pulses can give misleading relaxation data if spectral diffusion takes place \cite{Singer2002}, which is perhaps the case for \lsco (see below). Furthermore, large single crystals were not readily available and difficult to measure due to penetration depth effects. So, most early measurements were performed on ($c$-axis aligned) micro-crystalline powders with NMR or NQR (nuclear quadrupole resonance), which is the reason that most studies focussed on $1/T_{1\parallel}$, the rate measured if the crystal $c$-axis is parallel to the external field, which is also measured with NQR. 
The strongly underdoped systems were not investigated very much.
This is due to the fact that optimally doped materials have been of greatest interest, but also since underdoped cuprates show Cu signal wipe-out \cite{Hunt2001}, and one cannot assure that the measured signal represents the average material.

Nevertheless, while this all hampered rapid progress with NMR of cuprates, the nuclear relaxation data can be considered quite reliable for most systems.

In early NMR measurements on nearly optimally doped \ybco{} ($y\approx 0.95$) one had to assign the two different Cu sites, in the chains and plains, to two sets of NMR signals. 
Walstedt et al. in 1987 \cite{Walstedt1987} showed that one Cu site exhibited Fermi liquid-like relaxation above \tc, while the other showed significant deviations at higher temperatures. Then, since yttrium (Y) atoms that are sandwiched between the two \ce{CuO2} planes in this double-layer material, and since Markert et al. \cite{Markert1987} had reported Fermi liquid-like relaxation above \tc for Y, it was reasonable to assume \cite{Walstedt1987} that the Cu nuclei with Fermi liquid relaxation must be located in the plane. 
However, later it was shown from various experiments\cite{Kitaoka1988,Shimizu1988} that the opposite assignment was correct, which was put forward early on by Mali et al. \cite{Mali1987}. 
Later, Walstedt et al. \cite{Walstedt1988} discovered a nearly temperature independent anisotropy of relaxation ($T_{1\parallel}/T_{1\perp} \approx 3.4$) for planar Cu in \ybcoF{} above and below \tc (which we will show to be a unique property of the cuprates, only the proportionality factor can be material dependent). More importantly, it was concluded \cite{Walstedt1990} that the Korringa ratio is violated by comparing to shift data, in the sense that the relaxation is enhanced by an order of magnitude for planar Cu (and a factor of about 2.8 for planar O), which was taken as proof for enhanced antiferromagnetic spin fluctuations (that tend to cancel at planar O if the electron spins are located at planar Cu).

Despite the fact that only a very limited number of systems was investigated\cite{Pennington1988, Imai1988, Pennington1989,Takigawa1989,Takigawa1989b, Walstedt1989, Barret1991}, numerous models were developed to understand the nuclear relaxation, which will not be reviewed here.
Note that the mysterious cancellation of the Cu NMR shift for one orientation of the magnetic field (\cpara) was explained by an accidental cancellation of the on-site and transferred hyperfine coefficients ($A_\parallel+4B \approx 0$). 
Later, when systems were discovered that had a substantial shift also for \cpara (as much as 30\% of that for \cperp), this explanation was not questioned widely, while the corresponding changes in the hyperfine scenario appear unrealistic given the ubiquitous chemistry of the \ce{CuO2} plane.
With more thorough investigations of more systems, the then prevailing explanation did not grow more solid \cite{Slichter2007, Haase2009,Meissner2011,Haase2012,Rybicki2015}, rather, the failure of the hyperfine scenario became apparent \cite{Haase2017}. 
This questions, at the same time, any quantitative discussion of nuclear relaxation, which hinges on the hyperfine scenario that can filter out certain wave vectors from fluctuating modes that are relevant to nuclear relaxation. \medskip

Here we will take a fresh look at all available planar Cu NMR relaxation data of hole-doped cuprates to establish what we think is a new, but reliable phenomenology that may change the way one views some properties of the cuprates. 


\begin{table}[h]
	\begin{center}
		\footnotesize
		\caption{Summary of materials used for this review \cite{Itoh1996,Auler1999,Walstedt1989,Zimmermann1991,Magishi1996,Fujiwara1991,Zheng1996,Gerashenko1999,Itoh2017,Magishi1995,Gippius1999,Tokunaga2000,Walstedt1991,Bogdanovich1993}. Materials are listed with reference (Ref.), the apparent doping level (dop.), the \tc, the relaxation anisotropy ($\alpha_{\rm ani}$), and a colored symbol that is used throughout the manuscript.}
		\label{tab:one}
		\includegraphics[width=0.99\linewidth]{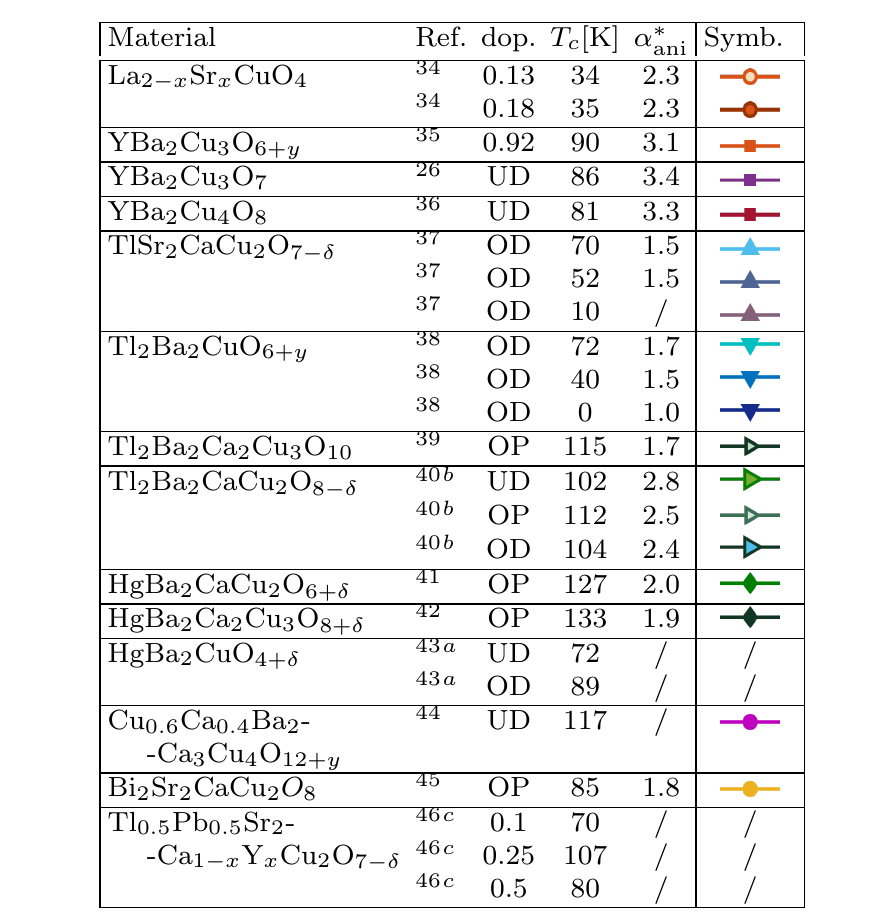}
		{\flushleft \raggedright\footnotesize These data are not considered in the analysis:\par
			$^{a}$ data contradictory to our own experimental data;\par
			$^{b}$ $T_{1\perp}$ not measured, but deduced from spin echo decay;\par
			$^{c}$ $T_{1}$ determination unclear and/or limited
			spectral resolution to measure site-specific $T_1$\\
			$^*$The error on $\alpha_{\rm ani}$ is typically less than $\pm 0.08$\\
		}
	\end{center}
\end{table}

By only plotting literature data we will establish that the relaxation in the cuprates is surprisingly simple and universal. 
It turns out that the large differences between different systems concern predominantly the relaxation measured with the external field along the crystal $c$-axis ($1/T_{1\parallel}$). However, since we also find that the relaxation anisotropy ($\alpha_{\rm ani}$) is \temp independent for all cuprates, i.e.,  $\alpha_\mathrm{ani} = T_{1\perp}/T_{1\parallel} $ above and below \tc, it is predominantly $\alpha_{\rm ani}$ that changes between different systems, i.e., it is the anisotropy of coupling to the electronic reservoir that varies, not the reservoir itself. Other than that, relaxation is material and doping independent and very similar to that for the most overdoped systems that are very close to Fermi liquids. Also below \tc the relaxation is very similar if plotted against $T/T_c$, the reduced temperature. Thus, there is no room for relaxation enhancement from spin fluctuations (except perhaps for the very underdoped systems for which we have no data).

\section{Results and discussion} \label{sR}

\begin{figure}[h!]
 \centering
 \includegraphics[width=0.6\linewidth]{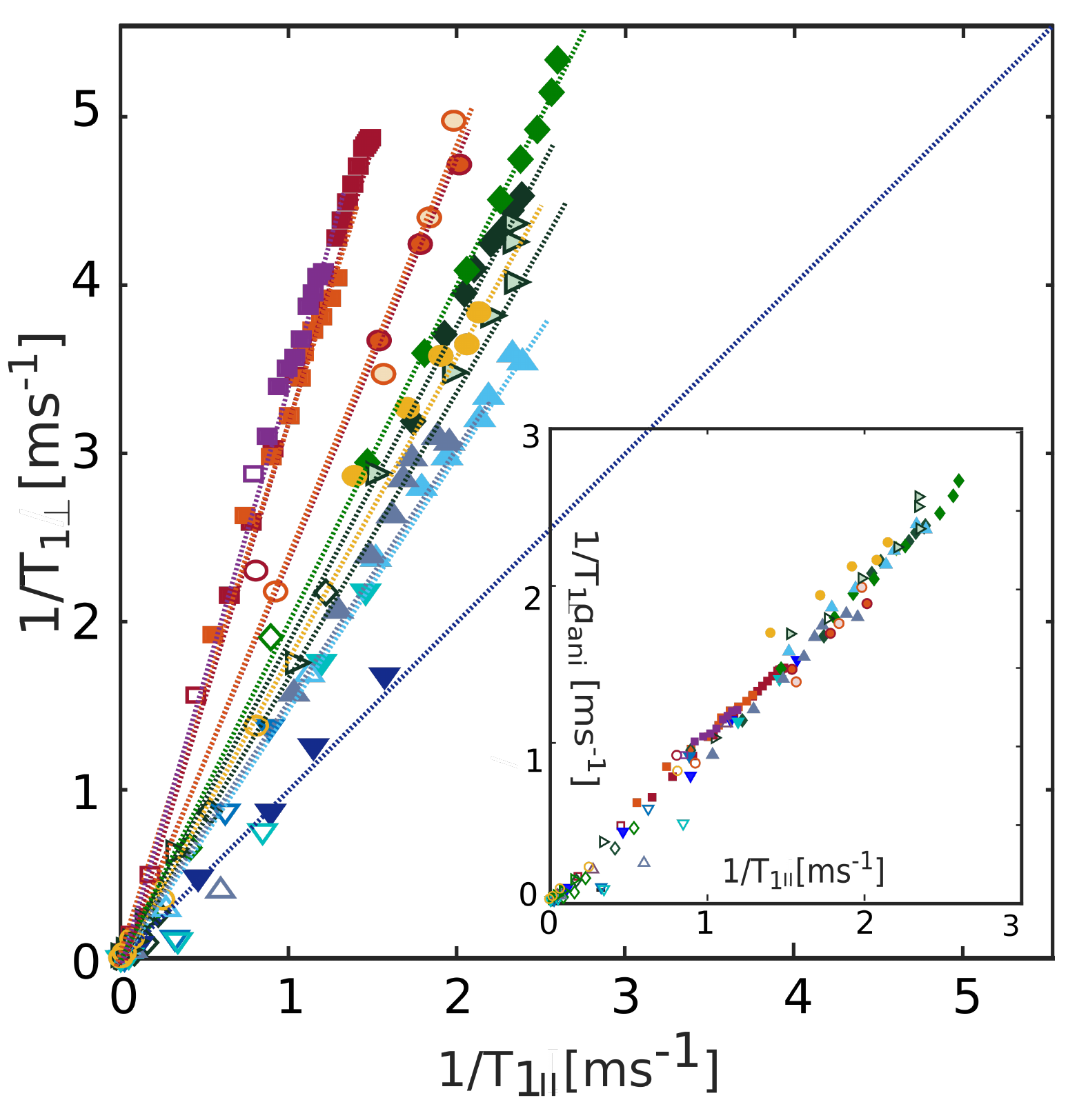}
 \caption{Main panel: $1/T_{1\perp}$ plotted vs. $1/T_{1\parallel}$ for each cuprate listed in Tab.~\ref{tab:one} (\temp is implicit parameter). Dotted lines are fits to the data for each cuprate with slopes given by Eq.~\eqref{eq:ani}. The blue dotted line is a diagonal  with slope $\alpha_{\rm ani}=1$. Inset: $1/T_{1\perp}\cdot 1/\alpha_{\rm ani}$ vs. $1/T_{1\parallel}$. All relaxation data can be explained by a single dominant relaxation process with a \temp-independent anisotropy.}
 \label{fig:three}
\end{figure}

In Tab.~\ref{tab:one} we list all cuprates, sorted by family, for which we could find data for both directions of the field. More information about data extraction and processing are given in the Appendix, together with a discussion of this representative selection of data.
Throughout the manuscript, data points are uniquely labelled as defined in Tab.~\ref{tab:one}. Furthermore, all displayed data points represent experimental data points from the literature, except for Fig.~\ref{fig:three} where we had to interpolate data points to be able to plot the data with \temp as an implicit parameter. 
A few sets of data are excluded from our discussion, nonetheless they are listed in Tab.~\ref{tab:one} (see Appendix).

Note that we assume magnetic, frequency-independent spin-lattice relaxation, as data were taken at different fields. This was proven a few times, but not for all data sets. We could not find significant differences between zero-field NQR and high-field $1/T_{1\parallel}$ data (for the same system). 

\begin{figure}[h]
 \centering
 \includegraphics[width=0.99\linewidth]{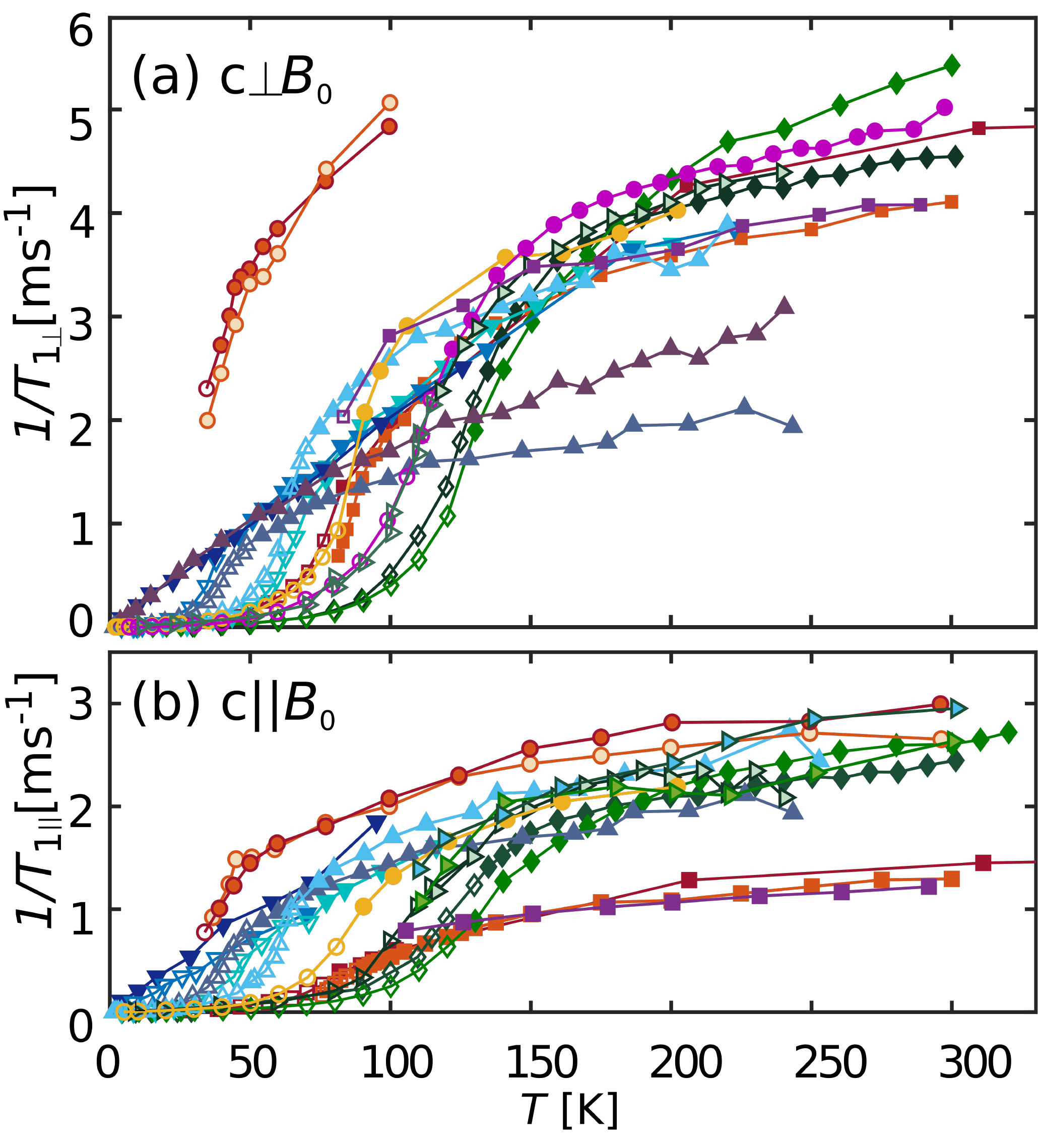}
 \caption{$^{63}$Cu nuclear relaxation rates as a function of temperature: (a) $1/T_{1\perp}$ and (b) $1/T_{1\parallel}$. For the used symbols see Tab.~\ref{tab:one}. Note that \textit{filled} symbols indicate data above $T_c$ and \textit{open} symbols data below $T_c$. For more explanations see text.}
\label{fig:two}
\end{figure}

 \subsubsection{General overview}
 An overview of all relaxation data as a function of temperature is given in Fig.~\ref{fig:two},  for two directions of the external field with respect to the crystal $c$-axis: (a), upper panel,  $1/T_{1\perp}$ for \cperp, and (b), lower panel, $1/T_{1\parallel}$ for \cpara. 

We would like to emphasize that it is more revealing to plot the relaxation rates against temperature (\temp), rather than plotting $1/(T_{1} T)$ as a function of \temp. For a Fermi liquid, $1/T_1 \propto T$, and both plots carry the same information. However, if other mechanisms are present they carry a \temp-dependent weight if one plots $1/(T_{1}T$), which complicates establishing a simple phenomenology.

While the plots are rather crowded one can conclude on some general behavior, already. (i) Except for a couple of outliers (that will be discussed later), both panels show similar dependences, i.e., starting from $T=0$ where relaxation has disappeared, the rates rise slowly below \tc. Above \tc, the rates are similar and begin to lag behind a Fermi liquid-behavior, eventually. (ii) The data in panel (b) for $1/T_{1\parallel}$, while considerably smaller than for the other direction, show greater variations for different materials, which is somewhat surprising as the anisotropy of the hyperfine coupling coefficients is not expected to change. (iii) For $1/T_{1\perp}$ (panel (a)) one can identify Fermi liquid-like relaxation for many systems, without or with a \tc. In fact, one could imagine that if one would suppress \tc, $1/T_{1\perp}$ is not far from Fermi liquid-behavior, with deviations mostly at higher temperatures (that can be caused by a more complicated band structure and has been observed in Fermi liquids \cite{Silbernagle1967}).
We will now discuss salient features observed in the data in more detail.
\subsubsection{Temperature independent $T_1$ anisotropy}
A salient feature in cuprate nuclear relaxation is a rather temperature independent anisotropy,
\begin{equation}\label{eq:ani}
\alpha_\mathrm{ani} = \frac{1/T_{1\perp}}{1/T_{1\parallel}} \equiv \frac{T_{1\parallel} }{T_{1\perp}}.
\end{equation}
That is, if we plot $1/T_{1\perp}$ vs. $1/T_{1\parallel}$ as in the main panel of Fig.~\ref{fig:three}, we find straight lines intersecting the origin, with slopes depending on the material. If one normalizes the slopes by the material specific $\alpha_{\rm ani}$, that is given in Tab.~\ref{tab:one}, all relaxation data collapse and fall on a single line as shown in the inset of Fig.~\ref{fig:three}. 

Interestingly, these slopes take on special values for various materials and/or doping levels, and $\alpha_{\rm ani}$ appears to increase as the doping decreases, but this is by no means a strict trend. The smallest  $\alpha_{\rm ani} = 1$ (isotropic behavior) is observed for the highest doping levels, and the largest of 3.4 for \ybcoF. For the underdoped, stoichiometric \ybcoE we find $\alpha_{\rm ani} = 3.33\pm0.02$, with rather high precision (this is one of the few cuprates that has very narrow linewidths). Indeed, it appears that $\alpha_{\rm ani}$ takes on special values rather than showing a smooth dependence. 

Since both rates are proportional to each other, above and below \tc, one concludes on a single, dominant relaxation mechanism with excitations that are present at the highest temperatures, and that decrease as \temp is lowered, similar to what happens for a Fermi liquid. Below \tc the changes are more rapid but still show a rather fixed $\alpha_{\rm ani}$ (note that the field's influence on \tc is anisotropic, as well, but those effects are mostly within the error bars here). 

One would argue that the electronic liquid behaves quite similar in all cuprates and that it is the anisotropy of the coupling of the nuclei to this electronic reservoir that changes with doping and material. Furthermore, if the anisotropic relaxation mechanism does not share the crystal symmetry exactly, the differences in panel (a) of Fig.~\ref{fig:two} could even be less.

Changes of the anisotropy of the hyperfine coefficients could lead to such behavior, but it is difficult to understand why this would produce only certain ratios and such large differences. While the spin response in the cuprates is believed to be rather isotropic (except for strongly underdoped systems), dynamic correlations on short length scales that can exist (as one knows from other probes) can contribute to such a behavior as well. 

\begin{figure}[h]
\centering
\includegraphics[width=0.6\linewidth]{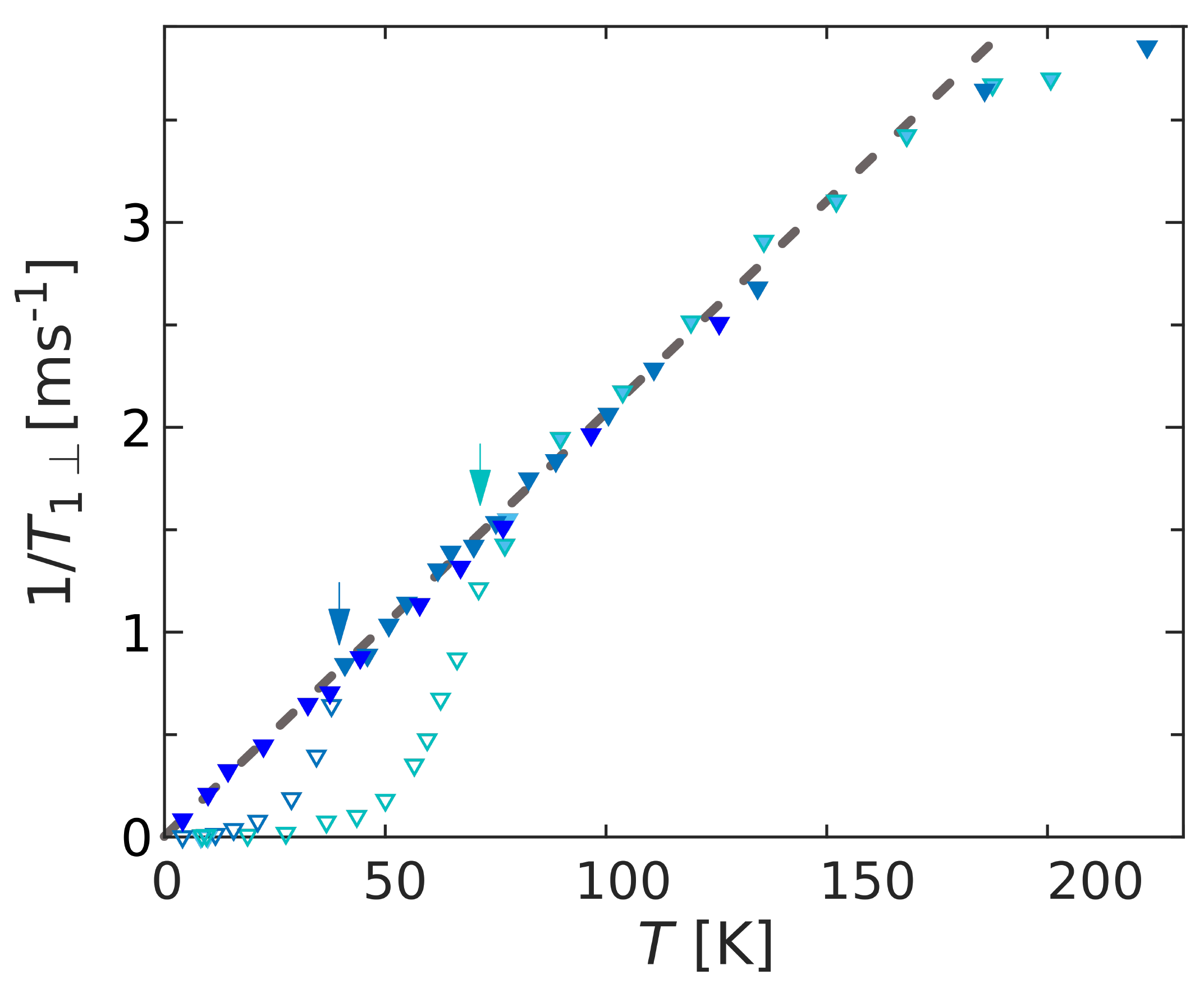}
\caption{Temperature dependence of the spin-lattice relaxation rates  $1/T_{1\perp}$ for differently overdoped Tl$_2$Ba$_2$CuO$_{6+y}$ materials with \tc = 0, 40, and 72 K (arrows). The grey dashed line has a slope of \SI{21}{/Ks} which follows from Korringa's law for the Knight shift of about 0.89\% for a simple Fermi liquid found for the highest doping level\cite{Haase2017}.}
 \label{fig:four}
\end{figure}
 
\subsubsection{Fermi liquid in overdoped Tl$_2$Ba$_2$CuO$_{6+y}$}
We now turn to the most overdoped cuprates in our data set, the Tl$_2$Ba$_2$CuO$_{6+y}$ family of materials. 
In Fig.~\ref{fig:four} we plot $1/T_{1\perp}$ as a function of temperature for different doping levels with \tc of 0, 40, and \SI{72}{K} (the same data are also present in Fig.~\ref{fig:two}). The dashed line is given by $1/T_{1\perp} = \sigma \cdot T$, with $\sigma=\SI{21}{/Ks}$ as for a Fermi liquid with a Knight shift of 0.89\% if one assumes $\rho =1$ in Korringa's formula, which is close to what has been measured\cite{Haase2017}.
As can already be seen in Fig.~\ref{fig:two} the data lag behind the Fermi liquid dependence only above about \SI{200}{K}.

Clearly, this is hallmark Fermi liquid behavior for the most overdoped system (below \SI{200}{K}). This is also true for the other two systems, except for a slight change in the anisotropies (there is no a priori reason to expect isotropic coupling). Below \tc, we observe spin singlet pairing without a significant enhancement from coherent scattering (Hebel-Slichter peak). Again, from these plots one would assume that these three systems are well-behaved Fermi liquids, at least below about \SI{200}{K}.

If one revisits Fig.~\ref{fig:two}, panel (a), with this important information one is forced to conclude that the cuprate relaxation behaves rather Fermi liquid-like below about \SI{200}{K} apart from the differences due to \tc for all doping levels and materials.

\subsubsection{Doping dependence of nuclear relaxation}
\begin{figure}[]
 \centering
 \includegraphics[width=1.0\linewidth]{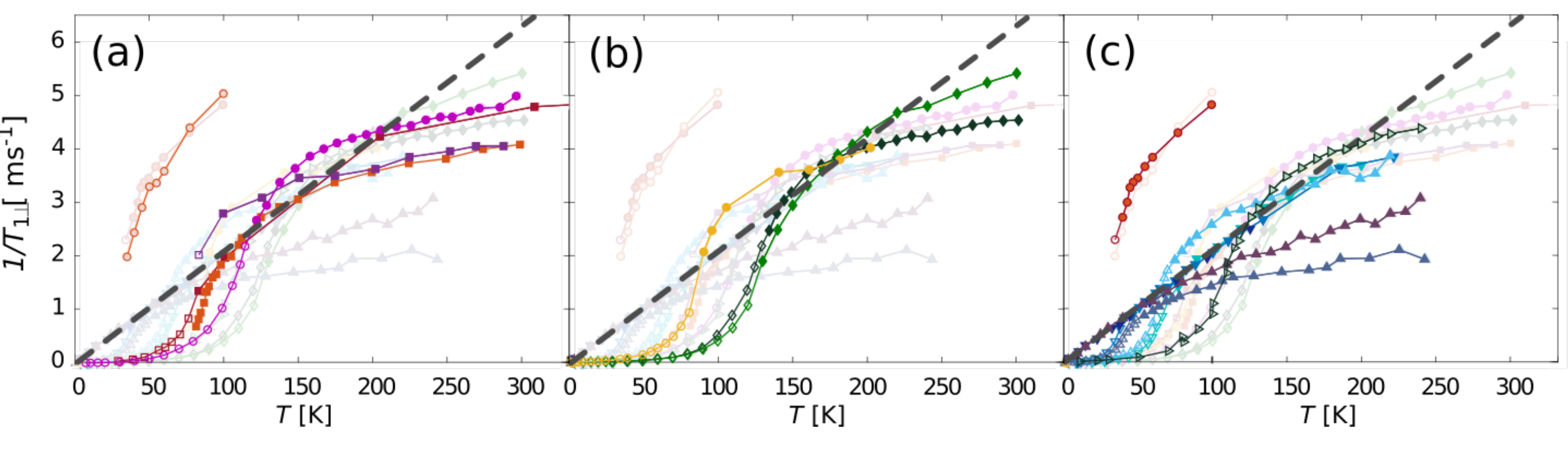}
 \caption{Relaxation rate $1/T_{1\perp}$ vs. \temp (the same Fig.~\ref{fig:two}), with emphasis to three different regions of the phase diagram: (a) underdoped, (b) optimally doped, and (c) overdoped materials. In addition, each panel shows the same dashed line according to a Fermi liquid with 0.89\% Knight shift.}
 \label{fig:five}
\end{figure}

In order to see how different doping levels affect the apparent relaxation we plot in Fig.~\ref{fig:five} the same data as in Fig.~\ref{fig:two}, but we emphasize in each of three panels a different doping range: (a) underdoped, (b) optimally doped, and (c) overdoped. Also shown is the dashed Fermi liquid line from Fig.~\ref{fig:four}.

Apart form the differences in \tc we do not see a particular trend in terms of doping dependence. It appears that no matter what the doping level is, whenever a material leaves the superconducting state, i.e., just above \tc, the relaxation is quite unique and very similar to the Fermi liquid value found for the very much overdoped systems. Also independent on doping, at higher temperatures the relaxation rate starts to lag behind the Fermi liquid temperature dependence. \lsco{} appears to be an outlier independent on doping, as well.

Note that we do not have data for the very underdoped materials so that the findings above may not be valid there.

\subsubsection{La$_{2-x}$Sr$_x$CuO$_4$}
A  significantly larger $1/T_{1\perp}$ compared to all other materials is found for the \lsco{} family. 
Such high rates \cite{Ohsugi1994} of $1/T_{1\perp}$ in \lsco{}  have been reported repeatedly. It was discussed that these rates show a doping-dependent paramagnetic contribution, i.e., $1/T_{1}(T)=const.$, as well as an antiferromagnetic contribution, $T_{1}T \propto (T+T_\mathrm{N})$.\cite{Gorkov2004} 

Also, it was reported that  $1/T_{1}$ converges at rather high \temp (above \SI{700}{K}) to a doping-independent value consistent with paramagnetic state of a Heisenberg antiferromagnet \cite{Imai1994}.

Given the additive nature of independent relaxation channels, it appears that \lsco{} may well have a similar component as all the other cuprates, i.e., one that is proportional to \temp, but also a second contribution that causes the special relaxation behavior. Also in terms of other (NMR) parameters \lsco{} appears to be somewhat of an outlier: local charges on planar O and Cu measured by NMR clearly show that it has by far the least covalent in-plane bonding, such that its inherent hole is almost entirely localized in Cu 3d$_{x^2-y^2}$ \cite{Jurkutat2014}.

In terms of Cu shift it also shows a special phenomenology, displaying no temperature or doping dependence of the shift for \cpara, and a comparatively strong dependence for \cperp \cite{Haase2017}.

\subsubsection{Relaxation below $T_c$}
\begin{figure}[]
 \centering
 \includegraphics[width=0.7\linewidth]{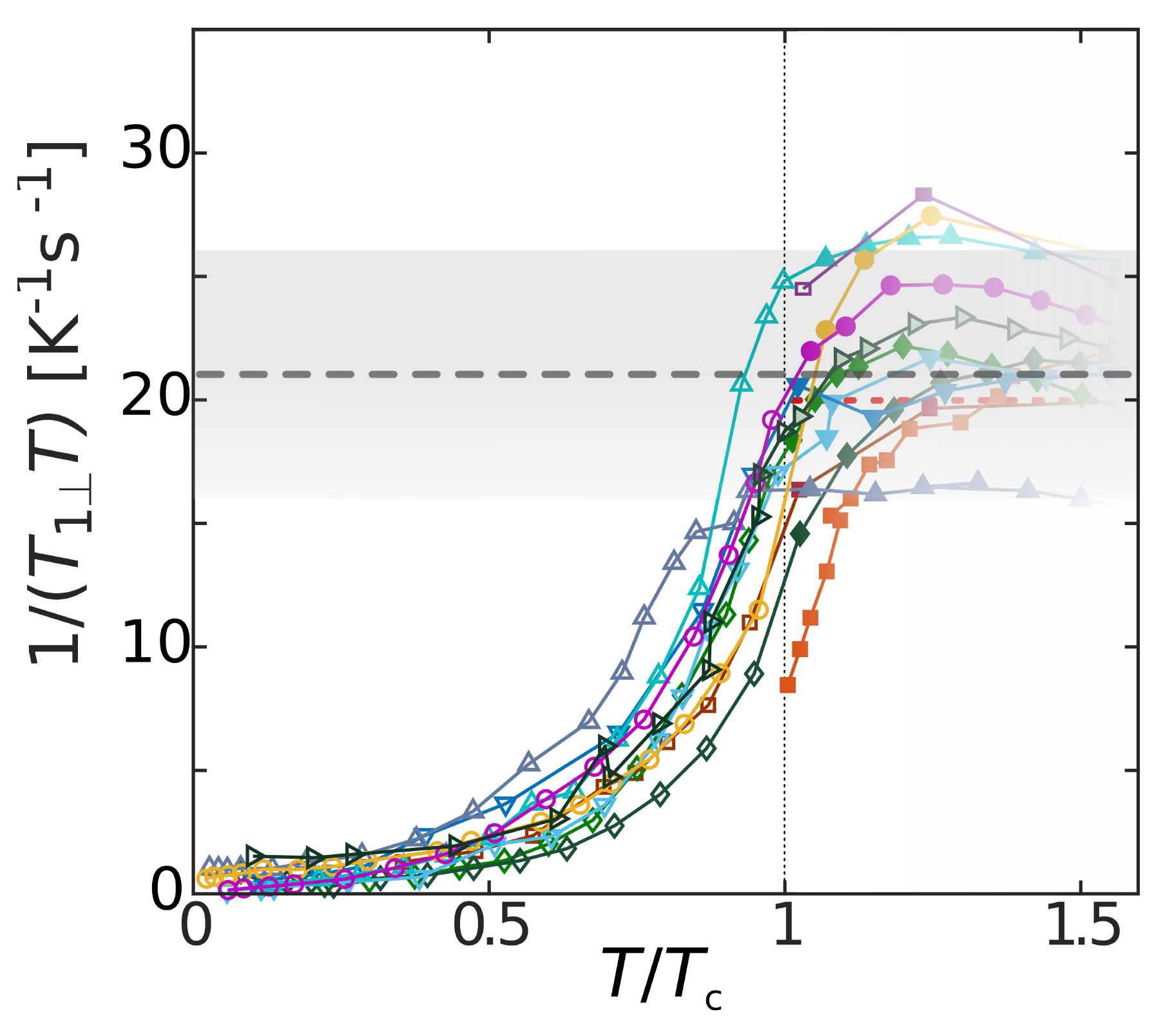}
 \caption{$(T_{1\perp}T)^{-1}$ in $c \perp B_0$ for all materials listed in Tab.~\ref{tab:one} as function of temperature scaled by the respective $T_c$. Also shown is $(T_1 T)^{-1} (T)=21 \pm5$~s$^{-1}$K$^{-1}$ (\textit{dashed gray line with shaded background}).}
 \label{fig:six}
\end{figure}

In order to see more clearly whether there is special behavior below \tc, we plot in Fig.~\ref{fig:six} the relaxation $1/(T_{1\perp}T)$ as a function of the reduced temperature $T/T_c$. We restrict the plot to $T/T_c \approx 1.5$ since one cannot expect the reduced \temp to be meaningful at higher \temp.

When one tries to evaluate this plot, one must keep in mind that the reported \tc, which is used for the scaling in Fig.~\ref{fig:six}, might not be the best choice for the actual, local energy scale (${\rm k_B}T_c$). For example, \tc could be suppressed by sample quality, or it might differ due to different definitions when measured with different techniques. Furthermore, any additional relaxation mechanism is scaled by \tc, as well, and may introduce differences in samples with very high vs. very low \tc. Finally, since we use only $1/T_{1\perp}$ a slight change in anisotropy could also affect this value. 

Despite possible uncertainties, inspection of Fig.~\ref{fig:six} shows rather unique behavior for $T < T_c$. Just above \tc almost all cuprates come up to a similar relaxation rate, as we recognized earlier. This points to the same mechanism in the superconducting region independent on material and doping. Also, within the small variation of dependences there is no clear trend as a function of the actual \tc or doping. Again, the most underdoped systems are almost indistinguishable from the most overdoped materials.

Worth mentioning is also that no cuprate shows an increased relaxation at the lowest temperatures in the $1/(T_{1}T)$, where the low-\temp rates are multiplied by increasingly large inverse \temp. So all excitations from this dominant mechanism are becoming gapped, which is true for $d$- and $s$-wave singlet pairing.

Whether the broad maximum in $1/(T_{1\perp}T) (T > T_c)$ seen in many materials signifies a vastly broadened coherence peak can unfortunately not be judged from the data available.

\section{Conclusions}
A review of all available planar Cu nuclear relaxation data in hole-doped cuprates offers a different understanding of the relevant electronic excitations.

The nuclear relaxation when the external field lies in the CuO$_2$ plane, $1/T_{1\perp}$, is found to be rather independent on family and doping -- from the weakly underdoped to even those very strongly overdoped systems that are close to an ideal Fermi liquid, for which Korringa's law holds and to which the nuclei couple isotropcially. 
The material dependent, and more often investigated $1/T_{1\parallel}$ is proportional to $1/T_{1\perp}$, above and below \tc, and thus only defines a material dependent anisotropy of the nuclear coupling to the electronic bath. 
Thus, the nuclei appear to experience rather ubiquitous electronic excitations that begin to freeze below \tc. Therefore, the bath itself appears Fermi liquid-like throughout the whole phase diagram for all systems. At higher temperatures the rates lag behind what is expected from a simple Fermi liquid (similar for all systems).

We also find universal behavior below \tc, i.e. the relaxation rates as a function of the reduced temperature ($T/T_c$) are rather similar.

All this points to a single, dominant relaxation mechanism due to electronic excitations that change significantly only below \tc due to spin singlet pairing. 

In particular, no special electronic spin fluctuations were found to enhance nuclear relaxation. Furthermore, the pseudogap does not seem to affect the Cu relaxation, while it was shown that it is important for the suppression of the NMR shifts \cite{Avramovska2018}. It was noticed before \cite{Berthier1997} that the Cu relaxation is in disagreement with neutron scattering results.

While we cannot say anything about the behavior of strongly underdoped systems, since there are no data available, it appear that only the \lsco family of materials is an outlier to the discussed scenario, as it appears to show additional relaxation for $1/T_{1\perp}$.\vspace{1cm}\par \medskip


\acknowledgements
We acknowledge the financial support from the University of Leipzig, the Free State of Saxony, the European Social Fund (ESF), and the Deutsche Forschungsgemeinschaft (DFG).\vspace{0.3cm}

{\flushleft Author} contributions: Data collection was performed equally by D.D and M.J.; independent verification of collected data was equally performed by M.A. and J.H.; discussion of data with equal help from D.P. and G.V.M.W.; preparation of the manuscript was equally performed by M.A., M.J., J.H.; J.H. also performed the data analysis and led the overall project.
\appendix

\section{Literature data processing} 

For the review of relaxation data we have collected all available literature data of $^{63}T_1$ of planar Cu. That means data for two orientations of the magnetic field with respect to the crystal $c$-axis, \cpara and \cperp, i.e. $1/T_{1\parallel}$ and $1/T_{1\perp}$, respectively. Furthermore, nuclear quadrupole resonance (NQR) were gathered, as well. The set comprises about 54 materials for $1/T_{1\parallel}$ .
The discussion in this manuscript, however, is limited to the 24 systems listed in Tab.~\ref{tab:one}, for which data for both directions of the field are available. Nevertheless, this (significant) subset we are discussing is representative of all the data in terms of amplitude and different temperature dependences of relaxation, as we can judge from all $1/T_{1\parallel}$ data.

As remarked in the main text, the higher abundance of $1/T_{1\parallel}$ data is due to the use of $c$-axis aligned powders and NQR.

We have excluded data on electron-doped cuprates where $1/T_1$ in most cases is affected by rare earth magnetism in the charge reservoir layer, data on antiferromagnetic inner layers in triple and higher layered materials, and data where it was unclear what definition for the $T_1$ was used\cite{Bogdanovich1993}.
We have also excluded HgBa$_2$CuO$_{4+\delta}$, for which our data are contradictory to results by Gippius et al. \cite{Gippius1999}, as well as Tl$_2$Ba$_2$CaCu$_2$O$_{8-\delta} $, since $^{63}T_{1 \perp}$ was not actually measured by Gerashenko et al. \cite{Gerashenko1999}, but deduced from the spin-echo decay.

The data were extracted using the online software "WebPlotDigitzier",  for which screenshots from graphs from the referenced papers were imported and the data extracted using the software tools. \\ 

In Fig.~\ref{fig:three} the temperature is an implicit parameter, owing to the limited availability of $1/T_{1\alpha} (T)$ data for both orientations at identical temperatures, we used a linear interpolation.

\bibliography{JHRelax}

\begin{thebibliography}{52}%
\makeatletter
\providecommand \@ifxundefined [1]{%
 \@ifx{#1\undefined}
}%
\providecommand \@ifnum [1]{%
 \ifnum #1\expandafter \@firstoftwo
 \else \expandafter \@secondoftwo
 \fi
}%
\providecommand \@ifx [1]{%
 \ifx #1\expandafter \@firstoftwo
 \else \expandafter \@secondoftwo
 \fi
}%
\providecommand \natexlab [1]{#1}%
\providecommand \enquote  [1]{``#1''}%
\providecommand \bibnamefont  [1]{#1}%
\providecommand \bibfnamefont [1]{#1}%
\providecommand \citenamefont [1]{#1}%
\providecommand \href@noop [0]{\@secondoftwo}%
\providecommand \href [0]{\begingroup \@sanitize@url \@href}%
\providecommand \@href[1]{\@@startlink{#1}\@@href}%
\providecommand \@@href[1]{\endgroup#1\@@endlink}%
\providecommand \@sanitize@url [0]{\catcode `\\12\catcode `\$12\catcode
  `\&12\catcode `\#12\catcode `\^12\catcode `\_12\catcode `\%12\relax}%
\providecommand \@@startlink[1]{}%
\providecommand \@@endlink[0]{}%
\providecommand \url  [0]{\begingroup\@sanitize@url \@url }%
\providecommand \@url [1]{\endgroup\@href {#1}{\urlprefix }}%
\providecommand \urlprefix  [0]{URL }%
\providecommand \Eprint [0]{\href }%
\providecommand \doibase [0]{http://dx.doi.org/}%
\providecommand \selectlanguage [0]{\@gobble}%
\providecommand \bibinfo  [0]{\@secondoftwo}%
\providecommand \bibfield  [0]{\@secondoftwo}%
\providecommand \translation [1]{[#1]}%
\providecommand \BibitemOpen [0]{}%
\providecommand \bibitemStop [0]{}%
\providecommand \bibitemNoStop [0]{.\EOS\space}%
\providecommand \EOS [0]{\spacefactor3000\relax}%
\providecommand \BibitemShut  [1]{\csname bibitem#1\endcsname}%
\let\auto@bib@innerbib\@empty
\bibitem [{\citenamefont {Slichter}(1990)}]{Slichter1990}%
  \BibitemOpen
  \bibfield  {author} {\bibinfo {author} {\bibfnamefont {C.~P.}\ \bibnamefont
  {Slichter}},\ }\href@noop {} {\emph {\bibinfo {title} {{Principles of
  Magnetic Resonance}}}},\ \bibinfo {edition} {3rd}\ ed.\ (\bibinfo
  {publisher} {Springer},\ \bibinfo {address} {Berlin},\ \bibinfo {year}
  {1990})\BibitemShut {NoStop}%
\bibitem [{\citenamefont {Heitler}\ and\ \citenamefont
  {Teller}(1936)}]{Heitler1936}%
  \BibitemOpen
  \bibfield  {author} {\bibinfo {author} {\bibfnamefont {W.}~\bibnamefont
  {Heitler}}\ and\ \bibinfo {author} {\bibfnamefont {E.}~\bibnamefont
  {Teller}},\ }\href {\doibase 10.1098/rspa.1936.0124} {\bibfield  {journal}
  {\bibinfo  {journal} {Proc. R. Soc. A Math. Phys. Eng. Sci.}\ }\textbf
  {\bibinfo {volume} {155}},\ \bibinfo {pages} {629} (\bibinfo {year}
  {1936})}\BibitemShut {NoStop}%
\bibitem [{\citenamefont {Korringa}(1950)}]{Korringa1950}%
  \BibitemOpen
  \bibfield  {author} {\bibinfo {author} {\bibfnamefont {J.}~\bibnamefont
  {Korringa}},\ }\href {\doibase 10.1016/0031-8914(50)90105-4} {\bibfield
  {journal} {\bibinfo  {journal} {Physica}\ }\textbf {\bibinfo {volume} {16}},\
  \bibinfo {pages} {601} (\bibinfo {year} {1950})}\BibitemShut {NoStop}%
\bibitem [{\citenamefont {Bardeen}\ \emph {et~al.}(1957)\citenamefont
  {Bardeen}, \citenamefont {Cooper},\ and\ \citenamefont
  {Schrieffer}}]{Bardeen1957}%
  \BibitemOpen
  \bibfield  {author} {\bibinfo {author} {\bibfnamefont {J.}~\bibnamefont
  {Bardeen}}, \bibinfo {author} {\bibfnamefont {L.~N.}\ \bibnamefont {Cooper}},
  \ and\ \bibinfo {author} {\bibfnamefont {J.~R.}\ \bibnamefont {Schrieffer}},\
  }\href {\doibase 10.1103/PhysRev.106.162} {\bibfield  {journal} {\bibinfo
  {journal} {Phys. Rev.}\ }\textbf {\bibinfo {volume} {106}},\ \bibinfo {pages}
  {162} (\bibinfo {year} {1957})}\BibitemShut {NoStop}%
\bibitem [{\citenamefont {Hebel}\ and\ \citenamefont
  {Slichter}(1959)}]{Hebel1957}%
  \BibitemOpen
  \bibfield  {author} {\bibinfo {author} {\bibfnamefont {L.~C.}\ \bibnamefont
  {Hebel}}\ and\ \bibinfo {author} {\bibfnamefont {C.~P.}\ \bibnamefont
  {Slichter}},\ }\href {\doibase 10.1103/PhysRev.113.1504} {\bibfield
  {journal} {\bibinfo  {journal} {Phys. Rev.}\ }\textbf {\bibinfo {volume}
  {113}},\ \bibinfo {pages} {1504} (\bibinfo {year} {1959})}\BibitemShut
  {NoStop}%
\bibitem [{\citenamefont {Bednorz}\ and\ \citenamefont
  {M{\"{u}}ller}(1986)}]{Bednorz1986}%
  \BibitemOpen
  \bibfield  {author} {\bibinfo {author} {\bibfnamefont {J.~G.}\ \bibnamefont
  {Bednorz}}\ and\ \bibinfo {author} {\bibfnamefont {K.~A.}\ \bibnamefont
  {M{\"{u}}ller}},\ }\href {\doibase 10.1007/BF01303701} {\bibfield  {journal}
  {\bibinfo  {journal} {Z. Phys. B Condens. Matter}\ }\textbf {\bibinfo
  {volume} {193}},\ \bibinfo {pages} {189} (\bibinfo {year}
  {1986})}\BibitemShut {NoStop}%
\bibitem [{\citenamefont {Kotegawa}\ \emph {et~al.}(2001)\citenamefont
  {Kotegawa}, \citenamefont {Ishida}, \citenamefont {Kitaoka}, \citenamefont
  {Muranaka},\ and\ \citenamefont {Akimitsu}}]{Kotegawa2001}%
  \BibitemOpen
  \bibfield  {author} {\bibinfo {author} {\bibfnamefont {H.}~\bibnamefont
  {Kotegawa}}, \bibinfo {author} {\bibfnamefont {K.}~\bibnamefont {Ishida}},
  \bibinfo {author} {\bibfnamefont {Y.}~\bibnamefont {Kitaoka}}, \bibinfo
  {author} {\bibfnamefont {T.}~\bibnamefont {Muranaka}}, \ and\ \bibinfo
  {author} {\bibfnamefont {J.}~\bibnamefont {Akimitsu}},\ }\href@noop {}
  {\bibfield  {journal} {\bibinfo  {journal} {Phys. Rev. Lett.}\ }\textbf
  {\bibinfo {volume} {87}},\ \bibinfo {pages} {127001} (\bibinfo {year}
  {2001})}\BibitemShut {NoStop}%
\bibitem [{\citenamefont {Silbernagle}\ \emph {et~al.}(1967)\citenamefont
  {Silbernagle}, \citenamefont {Weger}, \citenamefont {Clark},\ and\
  \citenamefont {Wernick}}]{Silbernagle1967}%
  \BibitemOpen
  \bibfield  {author} {\bibinfo {author} {\bibfnamefont {B.~G.}\ \bibnamefont
  {Silbernagle}}, \bibinfo {author} {\bibfnamefont {M.}~\bibnamefont {Weger}},
  \bibinfo {author} {\bibfnamefont {W.~G.}\ \bibnamefont {Clark}}, \ and\
  \bibinfo {author} {\bibfnamefont {J.~H.}\ \bibnamefont {Wernick}},\ }\href
  {\doibase 10.1103/PhysRev.153.535} {\bibfield  {journal} {\bibinfo  {journal}
  {Phys. Rev.}\ }\textbf {\bibinfo {volume} {153}},\ \bibinfo {pages} {535}
  (\bibinfo {year} {1967})}\BibitemShut {NoStop}%
\bibitem [{\citenamefont {Rybicki}\ \emph {et~al.}(2009)\citenamefont
  {Rybicki}, \citenamefont {Haase}, \citenamefont {Greven}, \citenamefont {Yu},
  \citenamefont {Li}, \citenamefont {Cho},\ and\ \citenamefont
  {Zhao}}]{Rybicki2009}%
  \BibitemOpen
  \bibfield  {author} {\bibinfo {author} {\bibfnamefont {D.}~\bibnamefont
  {Rybicki}}, \bibinfo {author} {\bibfnamefont {J.}~\bibnamefont {Haase}},
  \bibinfo {author} {\bibfnamefont {M.}~\bibnamefont {Greven}}, \bibinfo
  {author} {\bibfnamefont {G.}~\bibnamefont {Yu}}, \bibinfo {author}
  {\bibfnamefont {Y.}~\bibnamefont {Li}}, \bibinfo {author} {\bibfnamefont
  {Y.}~\bibnamefont {Cho}}, \ and\ \bibinfo {author} {\bibfnamefont
  {X.}~\bibnamefont {Zhao}},\ }\href {\doibase 10.1007/s10948-008-0376-2}
  {\bibfield  {journal} {\bibinfo  {journal} {J. Supercond. Nov. Magn.}\
  }\textbf {\bibinfo {volume} {22}},\ \bibinfo {pages} {179} (\bibinfo {year}
  {2009})}\BibitemShut {NoStop}%
\bibitem [{\citenamefont {Jurkutat}\ \emph {et~al.}(2013)\citenamefont
  {Jurkutat}, \citenamefont {Haase},\ and\ \citenamefont {Erb}}]{Jurkutat2013}%
  \BibitemOpen
  \bibfield  {author} {\bibinfo {author} {\bibfnamefont {M.}~\bibnamefont
  {Jurkutat}}, \bibinfo {author} {\bibfnamefont {J.}~\bibnamefont {Haase}}, \
  and\ \bibinfo {author} {\bibfnamefont {A.}~\bibnamefont {Erb}},\ }\href
  {\doibase 10.1007/s10948-013-2160-1} {\bibfield  {journal} {\bibinfo
  {journal} {J. Supercond. Nov. Magn.}\ }\textbf {\bibinfo {volume} {26}},\
  \bibinfo {pages} {2685} (\bibinfo {year} {2013})}\BibitemShut {NoStop}%
\bibitem [{\citenamefont {Reichardt}\ \emph {et~al.}(2018)\citenamefont
  {Reichardt}, \citenamefont {Jurkutat}, \citenamefont {Guehne}, \citenamefont
  {Kohlrautz}, \citenamefont {Erb},\ and\ \citenamefont
  {Haase}}]{Reichardt2018}%
  \BibitemOpen
  \bibfield  {author} {\bibinfo {author} {\bibfnamefont {S.}~\bibnamefont
  {Reichardt}}, \bibinfo {author} {\bibfnamefont {M.}~\bibnamefont {Jurkutat}},
  \bibinfo {author} {\bibfnamefont {R.}~\bibnamefont {Guehne}}, \bibinfo
  {author} {\bibfnamefont {J.}~\bibnamefont {Kohlrautz}}, \bibinfo {author}
  {\bibfnamefont {A.}~\bibnamefont {Erb}}, \ and\ \bibinfo {author}
  {\bibfnamefont {J.}~\bibnamefont {Haase}},\ }\href@noop {} {\bibfield
  {journal} {\bibinfo  {journal} {Condens. Matter}\ }\textbf {\bibinfo {volume}
  {3}},\ \bibinfo {pages} {23} (\bibinfo {year} {2018})}\BibitemShut {NoStop}%
\bibitem [{\citenamefont {Singer}\ \emph {et~al.}(2002)\citenamefont {Singer},
  \citenamefont {Hunt},\ and\ \citenamefont {Imai}}]{Singer2002}%
  \BibitemOpen
  \bibfield  {author} {\bibinfo {author} {\bibfnamefont {P.~M.}\ \bibnamefont
  {Singer}}, \bibinfo {author} {\bibfnamefont {A.~W.}\ \bibnamefont {Hunt}}, \
  and\ \bibinfo {author} {\bibfnamefont {T.}~\bibnamefont {Imai}},\ }\href
  {\doibase 10.1103/PhysRevLett.88.047602} {\bibfield  {journal} {\bibinfo
  {journal} {Phys. Rev. Lett.}\ }\textbf {\bibinfo {volume} {88}},\ \bibinfo
  {pages} {047602} (\bibinfo {year} {2002})}\BibitemShut {NoStop}%
\bibitem [{\citenamefont {Hunt}\ \emph {et~al.}(2001)\citenamefont {Hunt},
  \citenamefont {Singer}, \citenamefont {Cederstr{\"{o}}m},\ and\ \citenamefont
  {Imai}}]{Hunt2001}%
  \BibitemOpen
  \bibfield  {author} {\bibinfo {author} {\bibfnamefont {A.~W.}\ \bibnamefont
  {Hunt}}, \bibinfo {author} {\bibfnamefont {P.~P.~M.}\ \bibnamefont {Singer}},
  \bibinfo {author} {\bibfnamefont {A.~F.}\ \bibnamefont {Cederstr{\"{o}}m}}, \
  and\ \bibinfo {author} {\bibfnamefont {T.}~\bibnamefont {Imai}},\ }\href
  {\doibase 10.1103/PhysRevB.64.134525} {\bibfield  {journal} {\bibinfo
  {journal} {Phys. Rev. B}\ }\textbf {\bibinfo {volume} {64}},\ \bibinfo
  {pages} {134525} (\bibinfo {year} {2001})}\BibitemShut {NoStop}%
\bibitem [{\citenamefont {Walstedt}\ \emph {et~al.}(1987)\citenamefont
  {Walstedt}, \citenamefont {Warren}, \citenamefont {Bell}, \citenamefont
  {Brennert}, \citenamefont {Espinosa}, \citenamefont {Remeika}, \citenamefont
  {Cava},\ and\ \citenamefont {Rietman}}]{Walstedt1987}%
  \BibitemOpen
  \bibfield  {author} {\bibinfo {author} {\bibfnamefont {R.~E.}\ \bibnamefont
  {Walstedt}}, \bibinfo {author} {\bibfnamefont {W.~W.}\ \bibnamefont
  {Warren}}, \bibinfo {author} {\bibfnamefont {R.~F.}\ \bibnamefont {Bell}},
  \bibinfo {author} {\bibfnamefont {G.~F.}\ \bibnamefont {Brennert}}, \bibinfo
  {author} {\bibfnamefont {G.~P.}\ \bibnamefont {Espinosa}}, \bibinfo {author}
  {\bibfnamefont {J.~P.}\ \bibnamefont {Remeika}}, \bibinfo {author}
  {\bibfnamefont {R.~J.}\ \bibnamefont {Cava}}, \ and\ \bibinfo {author}
  {\bibfnamefont {E.~A.}\ \bibnamefont {Rietman}},\ }\href@noop {} {\bibfield
  {journal} {\bibinfo  {journal} {Phys. Rev. B}\ }\textbf {\bibinfo {volume}
  {36}},\ \bibinfo {pages} {5727} (\bibinfo {year} {1987})}\BibitemShut
  {NoStop}%
\bibitem [{\citenamefont {Markert}\ \emph {et~al.}(1987)\citenamefont
  {Markert}, \citenamefont {Noh}, \citenamefont {Russek},\ and\ \citenamefont
  {Cotts}}]{Markert1987}%
  \BibitemOpen
  \bibfield  {author} {\bibinfo {author} {\bibfnamefont {J.~T.}\ \bibnamefont
  {Markert}}, \bibinfo {author} {\bibfnamefont {T.~W.}\ \bibnamefont {Noh}},
  \bibinfo {author} {\bibfnamefont {S.~E.}\ \bibnamefont {Russek}}, \ and\
  \bibinfo {author} {\bibfnamefont {R.~M.}\ \bibnamefont {Cotts}},\ }\href@noop
  {} {\bibfield  {journal} {\bibinfo  {journal} {Solid State Commun.}\ }\textbf
  {\bibinfo {volume} {63}},\ \bibinfo {pages} {847} (\bibinfo {year}
  {1987})}\BibitemShut {NoStop}%
\bibitem [{\citenamefont {Kitaoka}\ \emph {et~al.}(1988)\citenamefont
  {Kitaoka}, \citenamefont {Hiramatsu}, \citenamefont {Kondo},\ and\
  \citenamefont {Asayama}}]{Kitaoka1988}%
  \BibitemOpen
  \bibfield  {author} {\bibinfo {author} {\bibfnamefont {Y.}~\bibnamefont
  {Kitaoka}}, \bibinfo {author} {\bibfnamefont {S.}~\bibnamefont {Hiramatsu}},
  \bibinfo {author} {\bibfnamefont {T.}~\bibnamefont {Kondo}}, \ and\ \bibinfo
  {author} {\bibfnamefont {K.}~\bibnamefont {Asayama}},\ }\href@noop {}
  {\bibfield  {journal} {\bibinfo  {journal} {J. Phys. Soc. Jpn.}\ }\textbf
  {\bibinfo {volume} {57}},\ \bibinfo {pages} {30} (\bibinfo {year}
  {1988})}\BibitemShut {NoStop}%
\bibitem [{\citenamefont {Shimizu}\ \emph {et~al.}(1988)\citenamefont
  {Shimizu}, \citenamefont {Yasuoka}, \citenamefont {Imai}, \citenamefont
  {Tsuda}, \citenamefont {Takabatake}, \citenamefont {Nakazawa},\ and\
  \citenamefont {Ishikawa}}]{Shimizu1988}%
  \BibitemOpen
  \bibfield  {author} {\bibinfo {author} {\bibfnamefont {T.}~\bibnamefont
  {Shimizu}}, \bibinfo {author} {\bibfnamefont {H.}~\bibnamefont {Yasuoka}},
  \bibinfo {author} {\bibfnamefont {T.}~\bibnamefont {Imai}}, \bibinfo {author}
  {\bibfnamefont {T.}~\bibnamefont {Tsuda}}, \bibinfo {author} {\bibfnamefont
  {T.}~\bibnamefont {Takabatake}}, \bibinfo {author} {\bibfnamefont
  {Y.}~\bibnamefont {Nakazawa}}, \ and\ \bibinfo {author} {\bibfnamefont
  {M.}~\bibnamefont {Ishikawa}},\ }\href@noop {} {\bibfield  {journal}
  {\bibinfo  {journal} {J. Phys. Soc. Jpn.}\ }\textbf {\bibinfo {volume}
  {57}},\ \bibinfo {pages} {2494} (\bibinfo {year} {1988})}\BibitemShut
  {NoStop}%
\bibitem [{\citenamefont {Mali}\ \emph {et~al.}(1987)\citenamefont {Mali},
  \citenamefont {Brinkmann}, \citenamefont {Pauli}, \citenamefont {Roos},
  \citenamefont {mann},\ and\ \citenamefont {Hulliger}}]{Mali1987}%
  \BibitemOpen
  \bibfield  {author} {\bibinfo {author} {\bibfnamefont {M.}~\bibnamefont
  {Mali}}, \bibinfo {author} {\bibfnamefont {D.}~\bibnamefont {Brinkmann}},
  \bibinfo {author} {\bibfnamefont {L.}~\bibnamefont {Pauli}}, \bibinfo
  {author} {\bibfnamefont {J.}~\bibnamefont {Roos}}, \bibinfo {author}
  {\bibfnamefont {H.}~\bibnamefont {mann}}, \ and\ \bibinfo {author}
  {\bibfnamefont {J.}~\bibnamefont {Hulliger}},\ }\href@noop {} {\bibfield
  {journal} {\bibinfo  {journal} {Phys. Lett. A}\ }\textbf {\bibinfo {volume}
  {124}},\ \bibinfo {pages} {112} (\bibinfo {year} {1987})}\BibitemShut
  {NoStop}%
\bibitem [{\citenamefont {Walstedt}\ \emph {et~al.}(1988)\citenamefont
  {Walstedt}, \citenamefont {Warren}, \citenamefont {Bell}, \citenamefont
  {Brennert}, \citenamefont {Espinosa}, \citenamefont {Cava}, \citenamefont
  {Schneemeyer},\ and\ \citenamefont {Waszczak}}]{Walstedt1988}%
  \BibitemOpen
  \bibfield  {author} {\bibinfo {author} {\bibfnamefont {R.~E.}\ \bibnamefont
  {Walstedt}}, \bibinfo {author} {\bibfnamefont {W.~W.}\ \bibnamefont
  {Warren}}, \bibinfo {author} {\bibfnamefont {R.~F.}\ \bibnamefont {Bell}},
  \bibinfo {author} {\bibfnamefont {G.~F.}\ \bibnamefont {Brennert}}, \bibinfo
  {author} {\bibfnamefont {G.~P.}\ \bibnamefont {Espinosa}}, \bibinfo {author}
  {\bibfnamefont {R.~J.}\ \bibnamefont {Cava}}, \bibinfo {author}
  {\bibfnamefont {L.~F.}\ \bibnamefont {Schneemeyer}}, \ and\ \bibinfo {author}
  {\bibfnamefont {J.~V.}\ \bibnamefont {Waszczak}},\ }\href@noop {} {\bibfield
  {journal} {\bibinfo  {journal} {Phys. Rev. B}\ }\textbf {\bibinfo {volume}
  {38}},\ \bibinfo {pages} {9299} (\bibinfo {year} {1988})}\BibitemShut
  {NoStop}%
\bibitem [{\citenamefont {Walstedt}\ and\ \citenamefont
  {Warren}(1990)}]{Walstedt1990}%
  \BibitemOpen
  \bibfield  {author} {\bibinfo {author} {\bibfnamefont {R.~E.}\ \bibnamefont
  {Walstedt}}\ and\ \bibinfo {author} {\bibfnamefont {W.~W.}\ \bibnamefont
  {Warren}},\ }\href@noop {} {\bibfield  {journal} {\bibinfo  {journal} {Phys.
  B: Condens. Matter}\ }\textbf {\bibinfo {volume} {163}},\ \bibinfo {pages}
  {75} (\bibinfo {year} {1990})}\BibitemShut {NoStop}%
\bibitem [{\citenamefont {Pennington}\ \emph {et~al.}(1988)\citenamefont
  {Pennington}, \citenamefont {Durand}, \citenamefont {Zax}, \citenamefont
  {Slichter}, \citenamefont {Rice},\ and\ \citenamefont
  {Ginsberg}}]{Pennington1988}%
  \BibitemOpen
  \bibfield  {author} {\bibinfo {author} {\bibfnamefont {C.~H.}\ \bibnamefont
  {Pennington}}, \bibinfo {author} {\bibfnamefont {D.~J.}\ \bibnamefont
  {Durand}}, \bibinfo {author} {\bibfnamefont {D.~B.}\ \bibnamefont {Zax}},
  \bibinfo {author} {\bibfnamefont {C.~P.}\ \bibnamefont {Slichter}}, \bibinfo
  {author} {\bibfnamefont {J.~P.}\ \bibnamefont {Rice}}, \ and\ \bibinfo
  {author} {\bibfnamefont {D.~M.}\ \bibnamefont {Ginsberg}},\ }\href@noop {}
  {\bibfield  {journal} {\bibinfo  {journal} {Phys. Rev. B}\ }\textbf {\bibinfo
  {volume} {37}},\ \bibinfo {pages} {7944} (\bibinfo {year}
  {1988})}\BibitemShut {NoStop}%
\bibitem [{\citenamefont {Imai}\ \emph {et~al.}(1988)\citenamefont {Imai},
  \citenamefont {Shimizu}, \citenamefont {Yasuoka}, \citenamefont {Ueda},\ and\
  \citenamefont {Kosuge}}]{Imai1988}%
  \BibitemOpen
  \bibfield  {author} {\bibinfo {author} {\bibfnamefont {T.}~\bibnamefont
  {Imai}}, \bibinfo {author} {\bibfnamefont {T.}~\bibnamefont {Shimizu}},
  \bibinfo {author} {\bibfnamefont {H.}~\bibnamefont {Yasuoka}}, \bibinfo
  {author} {\bibfnamefont {Y.}~\bibnamefont {Ueda}}, \ and\ \bibinfo {author}
  {\bibfnamefont {K.}~\bibnamefont {Kosuge}},\ }\href@noop {} {\bibfield
  {journal} {\bibinfo  {journal} {J. Phys. Soc. Jpn.}\ }\textbf {\bibinfo
  {volume} {57}},\ \bibinfo {pages} {2280} (\bibinfo {year}
  {1988})}\BibitemShut {NoStop}%
\bibitem [{\citenamefont {Pennington}\ \emph {et~al.}(1989)\citenamefont
  {Pennington}, \citenamefont {Durand}, \citenamefont {Slichter}, \citenamefont
  {Rice}, \citenamefont {Bukowski},\ and\ \citenamefont
  {Ginsberg}}]{Pennington1989}%
  \BibitemOpen
  \bibfield  {author} {\bibinfo {author} {\bibfnamefont {C.~H.}\ \bibnamefont
  {Pennington}}, \bibinfo {author} {\bibfnamefont {D.~J.}\ \bibnamefont
  {Durand}}, \bibinfo {author} {\bibfnamefont {C.~P.}\ \bibnamefont
  {Slichter}}, \bibinfo {author} {\bibfnamefont {J.~P.}\ \bibnamefont {Rice}},
  \bibinfo {author} {\bibfnamefont {E.~D.}\ \bibnamefont {Bukowski}}, \ and\
  \bibinfo {author} {\bibfnamefont {D.~M.}\ \bibnamefont {Ginsberg}},\ }\href
  {\doibase 10.1103/PhysRevB.39.2902} {\bibfield  {journal} {\bibinfo
  {journal} {Phys. Rev. B}\ }\textbf {\bibinfo {volume} {39}},\ \bibinfo
  {pages} {2902} (\bibinfo {year} {1989})}\BibitemShut {NoStop}%
\bibitem [{\citenamefont {Takigawa}\ \emph
  {et~al.}(1989{\natexlab{a}})\citenamefont {Takigawa}, \citenamefont {Hammel},
  \citenamefont {Heffner},\ and\ \citenamefont {Fisk}}]{Takigawa1989}%
  \BibitemOpen
  \bibfield  {author} {\bibinfo {author} {\bibfnamefont {M.}~\bibnamefont
  {Takigawa}}, \bibinfo {author} {\bibfnamefont {P.~C.}\ \bibnamefont
  {Hammel}}, \bibinfo {author} {\bibfnamefont {R.~H.}\ \bibnamefont {Heffner}},
  \ and\ \bibinfo {author} {\bibfnamefont {Z.}~\bibnamefont {Fisk}},\ }\href
  {\doibase 10.1103/PhysRevB.39.7371} {\bibfield  {journal} {\bibinfo
  {journal} {Phys. Rev. B}\ }\textbf {\bibinfo {volume} {39}},\ \bibinfo
  {pages} {7371} (\bibinfo {year} {1989}{\natexlab{a}})}\BibitemShut {NoStop}%
\bibitem [{\citenamefont {Takigawa}\ \emph
  {et~al.}(1989{\natexlab{b}})\citenamefont {Takigawa}, \citenamefont {Hammel},
  \citenamefont {Heffner}, \citenamefont {Fisk}, \citenamefont {Smith},\ and\
  \citenamefont {Schwarz}}]{Takigawa1989b}%
  \BibitemOpen
  \bibfield  {author} {\bibinfo {author} {\bibfnamefont {M.}~\bibnamefont
  {Takigawa}}, \bibinfo {author} {\bibfnamefont {P.~C.}\ \bibnamefont
  {Hammel}}, \bibinfo {author} {\bibfnamefont {R.~H.}\ \bibnamefont {Heffner}},
  \bibinfo {author} {\bibfnamefont {Z.}~\bibnamefont {Fisk}}, \bibinfo {author}
  {\bibfnamefont {J.~L.}\ \bibnamefont {Smith}}, \ and\ \bibinfo {author}
  {\bibfnamefont {R.~B.}\ \bibnamefont {Schwarz}},\ }\href {\doibase
  10.1103/PhysRevB.39.300} {\bibfield  {journal} {\bibinfo  {journal} {Phys.
  Rev. B}\ }\textbf {\bibinfo {volume} {39}},\ \bibinfo {pages} {300} (\bibinfo
  {year} {1989}{\natexlab{b}})}\BibitemShut {NoStop}%
\bibitem [{\citenamefont {Walstedt}\ \emph {et~al.}(1989)\citenamefont
  {Walstedt}, \citenamefont {Warren}, \citenamefont {Bell},\ and\ \citenamefont
  {Espinosa}}]{Walstedt1989}%
  \BibitemOpen
  \bibfield  {author} {\bibinfo {author} {\bibfnamefont {R.~E.}\ \bibnamefont
  {Walstedt}}, \bibinfo {author} {\bibfnamefont {W.~W.}\ \bibnamefont
  {Warren}}, \bibinfo {author} {\bibfnamefont {R.~F.}\ \bibnamefont {Bell}}, \
  and\ \bibinfo {author} {\bibfnamefont {G.~P.}\ \bibnamefont {Espinosa}},\
  }\href@noop {} {\bibfield  {journal} {\bibinfo  {journal} {Phys. Rev. B}\
  }\textbf {\bibinfo {volume} {40}},\ \bibinfo {pages} {2572} (\bibinfo {year}
  {1989})}\BibitemShut {NoStop}%
\bibitem [{\citenamefont {Barrett}\ \emph {et~al.}(1991)\citenamefont
  {Barrett}, \citenamefont {Martindale}, \citenamefont {Durand}, \citenamefont
  {Pennington}, \citenamefont {Slichter}, \citenamefont {Friedmann},
  \citenamefont {Rice},\ and\ \citenamefont {Ginsberg}}]{Barret1991}%
  \BibitemOpen
  \bibfield  {author} {\bibinfo {author} {\bibfnamefont {S.~E.}\ \bibnamefont
  {Barrett}}, \bibinfo {author} {\bibfnamefont {J.~A.}\ \bibnamefont
  {Martindale}}, \bibinfo {author} {\bibfnamefont {D.~J.}\ \bibnamefont
  {Durand}}, \bibinfo {author} {\bibfnamefont {C.~H.}\ \bibnamefont
  {Pennington}}, \bibinfo {author} {\bibfnamefont {C.~P.}\ \bibnamefont
  {Slichter}}, \bibinfo {author} {\bibfnamefont {T.~A.}\ \bibnamefont
  {Friedmann}}, \bibinfo {author} {\bibfnamefont {J.~P.}\ \bibnamefont {Rice}},
  \ and\ \bibinfo {author} {\bibfnamefont {D.~M.}\ \bibnamefont {Ginsberg}},\
  }\href {\doibase 10.1103/PhysRevLett.66.108} {\bibfield  {journal} {\bibinfo
  {journal} {Phys. Rev. Lett.}\ }\textbf {\bibinfo {volume} {66}},\ \bibinfo
  {pages} {108} (\bibinfo {year} {1991})}\BibitemShut {NoStop}%
\bibitem [{\citenamefont {Slichter}(2007)}]{Slichter2007}%
  \BibitemOpen
  \bibfield  {author} {\bibinfo {author} {\bibfnamefont {C.~P.}\ \bibnamefont
  {Slichter}},\ }in\ \href {\doibase 10.1007/978-0-387-68734-6\_5} {\emph
  {\bibinfo {booktitle} {Handbook of High-Temperature Superconductivity}}},\
  \bibinfo {editor} {edited by\ \bibinfo {editor} {\bibfnamefont {J.~R.}\
  \bibnamefont {Schrieffer}}\ and\ \bibinfo {editor} {\bibfnamefont {J.~S.}\
  \bibnamefont {Brooks}}}\ (\bibinfo  {publisher} {Springer},\ \bibinfo
  {address} {New York},\ \bibinfo {year} {2007})\ pp.\ \bibinfo {pages}
  {215--256}\BibitemShut {NoStop}%
\bibitem [{\citenamefont {Haase}\ \emph {et~al.}(2009)\citenamefont {Haase},
  \citenamefont {Goh}, \citenamefont {Meissner}, \citenamefont {Alireza},\ and\
  \citenamefont {Rybicki}}]{Haase2009}%
  \BibitemOpen
  \bibfield  {author} {\bibinfo {author} {\bibfnamefont {J.}~\bibnamefont
  {Haase}}, \bibinfo {author} {\bibfnamefont {S.~K.}\ \bibnamefont {Goh}},
  \bibinfo {author} {\bibfnamefont {T.}~\bibnamefont {Meissner}}, \bibinfo
  {author} {\bibfnamefont {P.~L.}\ \bibnamefont {Alireza}}, \ and\ \bibinfo
  {author} {\bibfnamefont {D.}~\bibnamefont {Rybicki}},\ }\href@noop {}
  {\bibfield  {journal} {\bibinfo  {journal} {Rev. Sci. Instrum.}\ }\textbf
  {\bibinfo {volume} {80}},\ \bibinfo {pages} {073905} (\bibinfo {year}
  {2009})}\BibitemShut {NoStop}%
\bibitem [{\citenamefont {Meissner}\ \emph {et~al.}(2011)\citenamefont
  {Meissner}, \citenamefont {Goh}, \citenamefont {Haase}, \citenamefont
  {Williams},\ and\ \citenamefont {Littlewood}}]{Meissner2011}%
  \BibitemOpen
  \bibfield  {author} {\bibinfo {author} {\bibfnamefont {T.}~\bibnamefont
  {Meissner}}, \bibinfo {author} {\bibfnamefont {S.~K.}\ \bibnamefont {Goh}},
  \bibinfo {author} {\bibfnamefont {J.}~\bibnamefont {Haase}}, \bibinfo
  {author} {\bibfnamefont {G.~V.~M.}\ \bibnamefont {Williams}}, \ and\ \bibinfo
  {author} {\bibfnamefont {P.~B.}\ \bibnamefont {Littlewood}},\ }\href
  {\doibase 10.1103/PhysRevB.83.220517} {\bibfield  {journal} {\bibinfo
  {journal} {Phys. Rev. B}\ }\textbf {\bibinfo {volume} {83}},\ \bibinfo
  {pages} {220517(R)} (\bibinfo {year} {2011})}\BibitemShut {NoStop}%
\bibitem [{\citenamefont {Haase}\ \emph {et~al.}(2012)\citenamefont {Haase},
  \citenamefont {Rybicki}, \citenamefont {Slichter}, \citenamefont {Greven},
  \citenamefont {Yu}, \citenamefont {Li},\ and\ \citenamefont
  {Zhao}}]{Haase2012}%
  \BibitemOpen
  \bibfield  {author} {\bibinfo {author} {\bibfnamefont {J.}~\bibnamefont
  {Haase}}, \bibinfo {author} {\bibfnamefont {D.}~\bibnamefont {Rybicki}},
  \bibinfo {author} {\bibfnamefont {C.~P.}\ \bibnamefont {Slichter}}, \bibinfo
  {author} {\bibfnamefont {M.}~\bibnamefont {Greven}}, \bibinfo {author}
  {\bibfnamefont {G.}~\bibnamefont {Yu}}, \bibinfo {author} {\bibfnamefont
  {Y.}~\bibnamefont {Li}}, \ and\ \bibinfo {author} {\bibfnamefont
  {X.}~\bibnamefont {Zhao}},\ }\href {\doibase 10.1103/PhysRevB.85.104517}
  {\bibfield  {journal} {\bibinfo  {journal} {Phys. Rev. B}\ }\textbf {\bibinfo
  {volume} {85}},\ \bibinfo {pages} {104517} (\bibinfo {year}
  {2012})}\BibitemShut {NoStop}%
\bibitem [{\citenamefont {Rybicki}\ \emph {et~al.}(2015)\citenamefont
  {Rybicki}, \citenamefont {Kohlrautz}, \citenamefont {Haase}, \citenamefont
  {Greven}, \citenamefont {Zhao}, \citenamefont {Chan}, \citenamefont {Dorow},\
  and\ \citenamefont {Veit}}]{Rybicki2015}%
  \BibitemOpen
  \bibfield  {author} {\bibinfo {author} {\bibfnamefont {D.}~\bibnamefont
  {Rybicki}}, \bibinfo {author} {\bibfnamefont {J.}~\bibnamefont {Kohlrautz}},
  \bibinfo {author} {\bibfnamefont {J.}~\bibnamefont {Haase}}, \bibinfo
  {author} {\bibfnamefont {M.}~\bibnamefont {Greven}}, \bibinfo {author}
  {\bibfnamefont {X.}~\bibnamefont {Zhao}}, \bibinfo {author} {\bibfnamefont
  {M.~K.}\ \bibnamefont {Chan}}, \bibinfo {author} {\bibfnamefont {C.~J.}\
  \bibnamefont {Dorow}}, \ and\ \bibinfo {author} {\bibfnamefont {M.~J.}\
  \bibnamefont {Veit}},\ }\href {\doibase 10.1103/PhysRevB.92.081115}
  {\bibfield  {journal} {\bibinfo  {journal} {Phys. Rev. B}\ }\textbf {\bibinfo
  {volume} {92}},\ \bibinfo {pages} {081115(R)} (\bibinfo {year}
  {2015})}\BibitemShut {NoStop}%
\bibitem [{\citenamefont {Haase}\ \emph {et~al.}(2017)\citenamefont {Haase},
  \citenamefont {Jurkutat},\ and\ \citenamefont {Kohlrautz}}]{Haase2017}%
  \BibitemOpen
  \bibfield  {author} {\bibinfo {author} {\bibfnamefont {J.}~\bibnamefont
  {Haase}}, \bibinfo {author} {\bibfnamefont {M.}~\bibnamefont {Jurkutat}}, \
  and\ \bibinfo {author} {\bibfnamefont {J.}~\bibnamefont {Kohlrautz}},\
  }\href@noop {} {\bibfield  {journal} {\bibinfo  {journal} {Condens. Matter}\
  }\textbf {\bibinfo {volume} {2}},\ \bibinfo {pages} {16} (\bibinfo {year}
  {2017})}\BibitemShut {NoStop}%
\bibitem [{\citenamefont {Itoh}\ \emph {et~al.}(1996)\citenamefont {Itoh},
  \citenamefont {Machi}, \citenamefont {Fukuoka}, \citenamefont {Tanabe},\ and\
  \citenamefont {Yasuoka}}]{Itoh1996}%
  \BibitemOpen
  \bibfield  {author} {\bibinfo {author} {\bibfnamefont {Y.}~\bibnamefont
  {Itoh}}, \bibinfo {author} {\bibfnamefont {T.}~\bibnamefont {Machi}},
  \bibinfo {author} {\bibfnamefont {A.}~\bibnamefont {Fukuoka}}, \bibinfo
  {author} {\bibfnamefont {K.}~\bibnamefont {Tanabe}}, \ and\ \bibinfo {author}
  {\bibfnamefont {H.}~\bibnamefont {Yasuoka}},\ }\href@noop {} {\bibfield
  {journal} {\bibinfo  {journal} {J. Phys. Soc. Jpn.}\ }\textbf {\bibinfo
  {volume} {65}},\ \bibinfo {pages} {3751} (\bibinfo {year}
  {1996})}\BibitemShut {NoStop}%
\bibitem [{\citenamefont {Auler}\ \emph {et~al.}(1999)\citenamefont {Auler},
  \citenamefont {Horvati{\'c}}, \citenamefont {Gillet}, \citenamefont
  {Berthier}, \citenamefont {Berthier}, \citenamefont {Carretta}, \citenamefont
  {Kitaoka}, \citenamefont {S{\'e}gransan},\ and\ \citenamefont
  {Henry}}]{Auler1999}%
  \BibitemOpen
  \bibfield  {author} {\bibinfo {author} {\bibfnamefont {T.}~\bibnamefont
  {Auler}}, \bibinfo {author} {\bibfnamefont {M.}~\bibnamefont {Horvati{\'c}}},
  \bibinfo {author} {\bibfnamefont {J.~A.}\ \bibnamefont {Gillet}}, \bibinfo
  {author} {\bibfnamefont {C.}~\bibnamefont {Berthier}}, \bibinfo {author}
  {\bibfnamefont {Y.}~\bibnamefont {Berthier}}, \bibinfo {author}
  {\bibfnamefont {P.}~\bibnamefont {Carretta}}, \bibinfo {author}
  {\bibfnamefont {Y.}~\bibnamefont {Kitaoka}}, \bibinfo {author} {\bibfnamefont
  {P.}~\bibnamefont {S{\'e}gransan}}, \ and\ \bibinfo {author} {\bibfnamefont
  {J.~Y.}\ \bibnamefont {Henry}},\ }\href@noop {} {\bibfield  {journal}
  {\bibinfo  {journal} {Phys. C: Supercond.}\ }\textbf {\bibinfo {volume}
  {313}},\ \bibinfo {pages} {255} (\bibinfo {year} {1999})}\BibitemShut
  {NoStop}%
\bibitem [{\citenamefont {Zimmermann}\ \emph {et~al.}(1991)\citenamefont
  {Zimmermann}, \citenamefont {Mali}, \citenamefont {Bankay},\ and\
  \citenamefont {Brinkmann}}]{Zimmermann1991}%
  \BibitemOpen
  \bibfield  {author} {\bibinfo {author} {\bibfnamefont {H.}~\bibnamefont
  {Zimmermann}}, \bibinfo {author} {\bibfnamefont {M.}~\bibnamefont {Mali}},
  \bibinfo {author} {\bibfnamefont {M.}~\bibnamefont {Bankay}}, \ and\ \bibinfo
  {author} {\bibfnamefont {D.}~\bibnamefont {Brinkmann}},\ }\href@noop {}
  {\bibfield  {journal} {\bibinfo  {journal} {Phys. C: Supercond.}\ }\textbf
  {\bibinfo {volume} {185-189}},\ \bibinfo {pages} {1145} (\bibinfo {year}
  {1991})}\BibitemShut {NoStop}%
\bibitem [{\citenamefont {Magishi}\ \emph {et~al.}(1996)\citenamefont
  {Magishi}, \citenamefont {Kitaoka}, \citenamefont {Zheng}, \citenamefont
  {Asayama}, \citenamefont {Kondo}, \citenamefont {Shimakawa}, \citenamefont
  {Manako},\ and\ \citenamefont {Kubo}}]{Magishi1996}%
  \BibitemOpen
  \bibfield  {author} {\bibinfo {author} {\bibfnamefont {K.}~\bibnamefont
  {Magishi}}, \bibinfo {author} {\bibfnamefont {Y.}~\bibnamefont {Kitaoka}},
  \bibinfo {author} {\bibfnamefont {G.-q.}\ \bibnamefont {Zheng}}, \bibinfo
  {author} {\bibfnamefont {K.}~\bibnamefont {Asayama}}, \bibinfo {author}
  {\bibfnamefont {T.}~\bibnamefont {Kondo}}, \bibinfo {author} {\bibfnamefont
  {Y.}~\bibnamefont {Shimakawa}}, \bibinfo {author} {\bibfnamefont
  {T.}~\bibnamefont {Manako}}, \ and\ \bibinfo {author} {\bibfnamefont
  {Y.}~\bibnamefont {Kubo}},\ }\href@noop {} {\bibfield  {journal} {\bibinfo
  {journal} {Phys. Rev. B}\ }\textbf {\bibinfo {volume} {54}},\ \bibinfo
  {pages} {10131} (\bibinfo {year} {1996})}\BibitemShut {NoStop}%
\bibitem [{\citenamefont {Fujiwara}\ \emph {et~al.}(1991)\citenamefont
  {Fujiwara}, \citenamefont {Kitaoka}, \citenamefont {Ishida}, \citenamefont
  {Asayama}, \citenamefont {Shimakawa}, \citenamefont {Manako},\ and\
  \citenamefont {Kubo}}]{Fujiwara1991}%
  \BibitemOpen
  \bibfield  {author} {\bibinfo {author} {\bibfnamefont {K.}~\bibnamefont
  {Fujiwara}}, \bibinfo {author} {\bibfnamefont {Y.}~\bibnamefont {Kitaoka}},
  \bibinfo {author} {\bibfnamefont {K.}~\bibnamefont {Ishida}}, \bibinfo
  {author} {\bibfnamefont {K.}~\bibnamefont {Asayama}}, \bibinfo {author}
  {\bibfnamefont {Y.}~\bibnamefont {Shimakawa}}, \bibinfo {author}
  {\bibfnamefont {T.}~\bibnamefont {Manako}}, \ and\ \bibinfo {author}
  {\bibfnamefont {Y.}~\bibnamefont {Kubo}},\ }\href@noop {} {\bibfield
  {journal} {\bibinfo  {journal} {Phys. C: Supercond.}\ }\textbf {\bibinfo
  {volume} {184}},\ \bibinfo {pages} {207} (\bibinfo {year}
  {1991})}\BibitemShut {NoStop}%
\bibitem [{\citenamefont {Zheng}\ \emph {et~al.}(1996)\citenamefont {Zheng},
  \citenamefont {Kitaoka}, \citenamefont {Asayama}, \citenamefont {Hamada},
  \citenamefont {Yamauchi},\ and\ \citenamefont {Tanaka}}]{Zheng1996}%
  \BibitemOpen
  \bibfield  {author} {\bibinfo {author} {\bibfnamefont {G.-q.}\ \bibnamefont
  {Zheng}}, \bibinfo {author} {\bibfnamefont {Y.}~\bibnamefont {Kitaoka}},
  \bibinfo {author} {\bibfnamefont {K.}~\bibnamefont {Asayama}}, \bibinfo
  {author} {\bibfnamefont {K.}~\bibnamefont {Hamada}}, \bibinfo {author}
  {\bibfnamefont {H.}~\bibnamefont {Yamauchi}}, \ and\ \bibinfo {author}
  {\bibfnamefont {S.}~\bibnamefont {Tanaka}},\ }\href {\doibase
  10.1016/0921-4534(96)00092-5} {\bibfield  {journal} {\bibinfo  {journal}
  {Phys. C: Supercond.}\ }\textbf {\bibinfo {volume} {260}},\ \bibinfo {pages}
  {197} (\bibinfo {year} {1996})}\BibitemShut {NoStop}%
\bibitem [{\citenamefont {Gerashenko}\ \emph {et~al.}(1999)\citenamefont
  {Gerashenko}, \citenamefont {Piskunov}, \citenamefont {Mikhalev},
  \citenamefont {Ananyev}, \citenamefont {Okulova}, \citenamefont
  {Verkhovskii}, \citenamefont {Yakubovskii}, \citenamefont {Shustov},\ and\
  \citenamefont {Trokiner}}]{Gerashenko1999}%
  \BibitemOpen
  \bibfield  {author} {\bibinfo {author} {\bibfnamefont {A.}~\bibnamefont
  {Gerashenko}}, \bibinfo {author} {\bibfnamefont {Y.}~\bibnamefont
  {Piskunov}}, \bibinfo {author} {\bibfnamefont {K.}~\bibnamefont {Mikhalev}},
  \bibinfo {author} {\bibfnamefont {A.}~\bibnamefont {Ananyev}}, \bibinfo
  {author} {\bibfnamefont {K.}~\bibnamefont {Okulova}}, \bibinfo {author}
  {\bibfnamefont {S.}~\bibnamefont {Verkhovskii}}, \bibinfo {author}
  {\bibfnamefont {A.}~\bibnamefont {Yakubovskii}}, \bibinfo {author}
  {\bibfnamefont {L.}~\bibnamefont {Shustov}}, \ and\ \bibinfo {author}
  {\bibfnamefont {A.}~\bibnamefont {Trokiner}},\ }\href {\doibase
  10.1016/S0921-4534(99)00528-6} {\bibfield  {journal} {\bibinfo  {journal}
  {Phys. C: Supercond.}\ }\textbf {\bibinfo {volume} {328}},\ \bibinfo {pages}
  {163} (\bibinfo {year} {1999})}\BibitemShut {NoStop}%
\bibitem [{\citenamefont {Itoh}\ \emph {et~al.}(2017)\citenamefont {Itoh},
  \citenamefont {Machi},\ and\ \citenamefont {Yamamoto}}]{Itoh2017}%
  \BibitemOpen
  \bibfield  {author} {\bibinfo {author} {\bibfnamefont {Y.}~\bibnamefont
  {Itoh}}, \bibinfo {author} {\bibfnamefont {T.}~\bibnamefont {Machi}}, \ and\
  \bibinfo {author} {\bibfnamefont {A.}~\bibnamefont {Yamamoto}},\ }\href@noop
  {} {\bibfield  {journal} {\bibinfo  {journal} {Phys. Rev. B}\ }\textbf
  {\bibinfo {volume} {95}},\ \bibinfo {pages} {094501} (\bibinfo {year}
  {2017})}\BibitemShut {NoStop}%
\bibitem [{\citenamefont {Magishi}\ \emph {et~al.}(1995)\citenamefont
  {Magishi}, \citenamefont {Kitaoka}, \citenamefont {Zheng}, \citenamefont
  {Asayama}, \citenamefont {Tokiwa}, \citenamefont {Iyo},\ and\ \citenamefont
  {Ihara}}]{Magishi1995}%
  \BibitemOpen
  \bibfield  {author} {\bibinfo {author} {\bibfnamefont {K.}~\bibnamefont
  {Magishi}}, \bibinfo {author} {\bibfnamefont {Y.}~\bibnamefont {Kitaoka}},
  \bibinfo {author} {\bibfnamefont {G.-q.}\ \bibnamefont {Zheng}}, \bibinfo
  {author} {\bibfnamefont {K.}~\bibnamefont {Asayama}}, \bibinfo {author}
  {\bibfnamefont {K.}~\bibnamefont {Tokiwa}}, \bibinfo {author} {\bibfnamefont
  {A.}~\bibnamefont {Iyo}}, \ and\ \bibinfo {author} {\bibfnamefont
  {H.}~\bibnamefont {Ihara}},\ }\href {\doibase 10.1143/JPSJ.64.4561}
  {\bibfield  {journal} {\bibinfo  {journal} {J. Phys. Soc. Japan}\ }\textbf
  {\bibinfo {volume} {64}},\ \bibinfo {pages} {4561} (\bibinfo {year}
  {1995})}\BibitemShut {NoStop}%
\bibitem [{\citenamefont {Gippius}\ \emph {et~al.}(1999)\citenamefont
  {Gippius}, \citenamefont {Antipov}, \citenamefont {Hoffmann}, \citenamefont
  {L{\"{u}}ders},\ and\ \citenamefont {Buntkowsky}}]{Gippius1999}%
  \BibitemOpen
  \bibfield  {author} {\bibinfo {author} {\bibfnamefont {A.~A.}\ \bibnamefont
  {Gippius}}, \bibinfo {author} {\bibfnamefont {E.~V.}\ \bibnamefont
  {Antipov}}, \bibinfo {author} {\bibfnamefont {W.}~\bibnamefont {Hoffmann}},
  \bibinfo {author} {\bibfnamefont {K.}~\bibnamefont {L{\"{u}}ders}}, \ and\
  \bibinfo {author} {\bibfnamefont {G.}~\bibnamefont {Buntkowsky}},\
  }\href@noop {} {\bibfield  {journal} {\bibinfo  {journal} {Phys. Rev. B}\
  }\textbf {\bibinfo {volume} {59}},\ \bibinfo {pages} {654} (\bibinfo {year}
  {1999})}\BibitemShut {NoStop}%
\bibitem [{\citenamefont {Tokunaga}\ \emph {et~al.}(2000)\citenamefont
  {Tokunaga}, \citenamefont {Ishida}, \citenamefont {Kitaoka}, \citenamefont
  {Asayama}, \citenamefont {Tokiwa}, \citenamefont {Iyo},\ and\ \citenamefont
  {Ihara}}]{Tokunaga2000}%
  \BibitemOpen
  \bibfield  {author} {\bibinfo {author} {\bibfnamefont {Y.}~\bibnamefont
  {Tokunaga}}, \bibinfo {author} {\bibfnamefont {K.}~\bibnamefont {Ishida}},
  \bibinfo {author} {\bibfnamefont {Y.}~\bibnamefont {Kitaoka}}, \bibinfo
  {author} {\bibfnamefont {K.}~\bibnamefont {Asayama}}, \bibinfo {author}
  {\bibfnamefont {K.}~\bibnamefont {Tokiwa}}, \bibinfo {author} {\bibfnamefont
  {A.}~\bibnamefont {Iyo}}, \ and\ \bibinfo {author} {\bibfnamefont
  {H.}~\bibnamefont {Ihara}},\ }\href {\doibase 10.1103/PhysRevB.61.9707}
  {\bibfield  {journal} {\bibinfo  {journal} {Phys. Rev. B}\ }\textbf {\bibinfo
  {volume} {61}},\ \bibinfo {pages} {9707} (\bibinfo {year}
  {2000})}\BibitemShut {NoStop}%
\bibitem [{\citenamefont {Walstedt}\ \emph {et~al.}(1991)\citenamefont
  {Walstedt}, \citenamefont {Bell},\ and\ \citenamefont
  {Mitzi}}]{Walstedt1991}%
  \BibitemOpen
  \bibfield  {author} {\bibinfo {author} {\bibfnamefont {R.~E.}\ \bibnamefont
  {Walstedt}}, \bibinfo {author} {\bibfnamefont {R.~F.}\ \bibnamefont {Bell}},
  \ and\ \bibinfo {author} {\bibfnamefont {D.~B.}\ \bibnamefont {Mitzi}},\
  }\href {\doibase 10.1103/PhysRevB.44.7760} {\bibfield  {journal} {\bibinfo
  {journal} {Phys. Rev. B}\ }\textbf {\bibinfo {volume} {44}},\ \bibinfo
  {pages} {7760} (\bibinfo {year} {1991})}\BibitemShut {NoStop}%
\bibitem [{\citenamefont {Bogdanovich}\ \emph {et~al.}(1993)\citenamefont
  {Bogdanovich}, \citenamefont {Zhdanov}, \citenamefont {Mikhalyov},
  \citenamefont {Lavrentjev}, \citenamefont {Aleksashin}, \citenamefont
  {Verkovskij}, \citenamefont {Winzek}, \citenamefont {Gergen}, \citenamefont
  {Gross}, \citenamefont {Mehring} \emph {et~al.}}]{Bogdanovich1993}%
  \BibitemOpen
  \bibfield  {author} {\bibinfo {author} {\bibfnamefont {A.}~\bibnamefont
  {Bogdanovich}}, \bibinfo {author} {\bibfnamefont {Y.~I.}\ \bibnamefont
  {Zhdanov}}, \bibinfo {author} {\bibfnamefont {K.}~\bibnamefont {Mikhalyov}},
  \bibinfo {author} {\bibfnamefont {V.}~\bibnamefont {Lavrentjev}}, \bibinfo
  {author} {\bibfnamefont {B.}~\bibnamefont {Aleksashin}}, \bibinfo {author}
  {\bibfnamefont {S.}~\bibnamefont {Verkovskij}}, \bibinfo {author}
  {\bibfnamefont {N.}~\bibnamefont {Winzek}}, \bibinfo {author} {\bibfnamefont
  {P.}~\bibnamefont {Gergen}}, \bibinfo {author} {\bibfnamefont
  {J.}~\bibnamefont {Gross}}, \bibinfo {author} {\bibfnamefont
  {M.}~\bibnamefont {Mehring}},  \emph {et~al.},\ }\href@noop {} {\bibfield
  {journal} {\bibinfo  {journal} {Phys. C: Supercond.}\ }\textbf {\bibinfo
  {volume} {215}},\ \bibinfo {pages} {253} (\bibinfo {year}
  {1993})}\BibitemShut {NoStop}%
\bibitem [{\citenamefont {Ohsugi}\ \emph {et~al.}(1994)\citenamefont {Ohsugi},
  \citenamefont {Kitaoka}, \citenamefont {Ishida}, \citenamefont {Zheng},\ and\
  \citenamefont {Asayama}}]{Ohsugi1994}%
  \BibitemOpen
  \bibfield  {author} {\bibinfo {author} {\bibfnamefont {S.}~\bibnamefont
  {Ohsugi}}, \bibinfo {author} {\bibfnamefont {Y.}~\bibnamefont {Kitaoka}},
  \bibinfo {author} {\bibfnamefont {K.}~\bibnamefont {Ishida}}, \bibinfo
  {author} {\bibfnamefont {G.-q.}\ \bibnamefont {Zheng}}, \ and\ \bibinfo
  {author} {\bibfnamefont {K.}~\bibnamefont {Asayama}},\ }\href {\doibase
  10.1143/JPSJ.63.700} {\bibfield  {journal} {\bibinfo  {journal} {J. Phys.
  Soc. Jpn.}\ }\textbf {\bibinfo {volume} {63}},\ \bibinfo {pages} {700}
  (\bibinfo {year} {1994})}\BibitemShut {NoStop}%
\bibitem [{\citenamefont {Gor'kov}\ and\ \citenamefont
  {Teitel'baum}(2004)}]{Gorkov2004}%
  \BibitemOpen
  \bibfield  {author} {\bibinfo {author} {\bibfnamefont {L.~P.}\ \bibnamefont
  {Gor'kov}}\ and\ \bibinfo {author} {\bibfnamefont {G.~B.}\ \bibnamefont
  {Teitel'baum}},\ }\href {\doibase 10.1134/1.1808849} {\bibfield  {journal}
  {\bibinfo  {journal} {J. Exp. Theor. Phys. Lett.}\ }\textbf {\bibinfo
  {volume} {80}},\ \bibinfo {pages} {195} (\bibinfo {year} {2004})}\BibitemShut
  {NoStop}%
\bibitem [{\citenamefont {Imai}\ \emph {et~al.}(1994)\citenamefont {Imai},
  \citenamefont {Slichter}, \citenamefont {Yoshimura}, \citenamefont {Katoh},\
  and\ \citenamefont {Kosuge}}]{Imai1994}%
  \BibitemOpen
  \bibfield  {author} {\bibinfo {author} {\bibfnamefont {T.}~\bibnamefont
  {Imai}}, \bibinfo {author} {\bibfnamefont {C.~P.}\ \bibnamefont {Slichter}},
  \bibinfo {author} {\bibfnamefont {K.}~\bibnamefont {Yoshimura}}, \bibinfo
  {author} {\bibfnamefont {M.}~\bibnamefont {Katoh}}, \ and\ \bibinfo {author}
  {\bibfnamefont {K.}~\bibnamefont {Kosuge}},\ }\href@noop {} {\bibfield
  {journal} {\bibinfo  {journal} {Phys. B: Condens. Matter}\ }\textbf {\bibinfo
  {volume} {197}},\ \bibinfo {pages} {601} (\bibinfo {year}
  {1994})}\BibitemShut {NoStop}%
\bibitem [{\citenamefont {Jurkutat}\ \emph {et~al.}(2014)\citenamefont
  {Jurkutat}, \citenamefont {Rybicki}, \citenamefont {Sushkov}, \citenamefont
  {Williams}, \citenamefont {Erb},\ and\ \citenamefont {Haase}}]{Jurkutat2014}%
  \BibitemOpen
  \bibfield  {author} {\bibinfo {author} {\bibfnamefont {M.}~\bibnamefont
  {Jurkutat}}, \bibinfo {author} {\bibfnamefont {D.}~\bibnamefont {Rybicki}},
  \bibinfo {author} {\bibfnamefont {O.~P.}\ \bibnamefont {Sushkov}}, \bibinfo
  {author} {\bibfnamefont {G.~V.~M.}\ \bibnamefont {Williams}}, \bibinfo
  {author} {\bibfnamefont {A.}~\bibnamefont {Erb}}, \ and\ \bibinfo {author}
  {\bibfnamefont {J.}~\bibnamefont {Haase}},\ }\href {\doibase
  10.1103/PhysRevB.90.140504} {\bibfield  {journal} {\bibinfo  {journal} {Phys.
  Rev. B}\ }\textbf {\bibinfo {volume} {90}},\ \bibinfo {pages} {140504(R)}
  (\bibinfo {year} {2014})}\BibitemShut {NoStop}%
\bibitem [{\citenamefont {Avramovska}\ \emph {et~al.}(2018)\citenamefont
  {Avramovska}, \citenamefont {Pavicevic},\ and\ \citenamefont
  {Haase}}]{Avramovska2018}%
  \BibitemOpen
  \bibfield  {author} {\bibinfo {author} {\bibfnamefont {M.}~\bibnamefont
  {Avramovska}}, \bibinfo {author} {\bibfnamefont {D.}~\bibnamefont
  {Pavicevic}}, \ and\ \bibinfo {author} {\bibfnamefont {J.}~\bibnamefont
  {Haase}},\ }\href@noop {} {\bibfield  {journal} {\bibinfo  {journal}
  {arXiv:1806.03644}\ } (\bibinfo {year} {2018})}\BibitemShut {NoStop}%
\bibitem [{\citenamefont {Berthier}\ \emph {et~al.}(1997)\citenamefont
  {Berthier}, \citenamefont {Julien}, \citenamefont {Bakharev}, \citenamefont
  {Horvati{\'c}},\ and\ \citenamefont {S{\'e}gransan}}]{Berthier1997}%
  \BibitemOpen
  \bibfield  {author} {\bibinfo {author} {\bibfnamefont {C.}~\bibnamefont
  {Berthier}}, \bibinfo {author} {\bibfnamefont {M.-H.}\ \bibnamefont
  {Julien}}, \bibinfo {author} {\bibfnamefont {O.}~\bibnamefont {Bakharev}},
  \bibinfo {author} {\bibfnamefont {M.}~\bibnamefont {Horvati{\'c}}}, \ and\
  \bibinfo {author} {\bibfnamefont {P.}~\bibnamefont {S{\'e}gransan}},\
  }\href@noop {} {\bibfield  {journal} {\bibinfo  {journal} {Physica C}\
  }\textbf {\bibinfo {volume} {282}},\ \bibinfo {pages} {227} (\bibinfo {year}
  {1997})}\BibitemShut {NoStop}%
\end{thebibliography}%


\end{document}